\documentclass[12pt,oneside,reqno,american]{amsart}
\usepackage{biolinum}
\usepackage[T1]{fontenc}
\usepackage[latin9]{inputenc}
\usepackage{color}
\definecolor{lyxboxbgcolor}{rgb}{1, 0, 0}
\definecolor{lyxnotefontcolor}{rgb}{0.75, 0.75, 0.75}
\usepackage{verbatim}
\usepackage{booktabs}
\usepackage{amsbsy}
\usepackage{amstext}
\usepackage{amsthm}
\usepackage{amssymb}
\usepackage{esint}
\usepackage{xargs}[2008/03/08]

\makeatletter

\providecommand{\tabularnewline}{\\}

\numberwithin{equation}{section}
\numberwithin{figure}{section}
\numberwithin{table}{section}

\usepackage{mathtools}
\usepackage{amsthm}
\usepackage{pdfsync} 
\usepackage{hyperref}
\usepackage[all]{xy}

\usepackage[T1]{fontenc}
\usepackage{Alegreya} 
\usepackage[stix2,scaled=0.96]{newtxmath}
\usepackage[cal=boondoxo,bb=boondox,frak=boondox,scaled=1.10]{mathalfa}
\usepackage[scaled=.9]{rsfso}
\usepackage{xcolor} 
\definecolor{brown(traditional)}{rgb}{0.59, 0.29, 0.0}
\definecolor{blue(ryb)}{rgb}{0.01, 0.28, 1.0}
\definecolor{red}{rgb}{1.0, 0.0, 0.0}
\definecolor{magenta}{rgb}{1.0, 0.0, 1.0}
\definecolor{mahogany}{rgb}{0.75, 0.25, 0.0}
\definecolor{lavenderpurple}{rgb}{0.59, 0.48, 0.71}
\definecolor{olive}{rgb}{0.5, 0.5, 0.0}
\definecolor{brickred}{rgb}{0.8, 0.25, 0.33}
\definecolor{antiquefuchsia}{rgb}{0.57, 0.36, 0.51}
\definecolor{bole}{rgb}{0.47, 0.27, 0.23}
\definecolor{darkolivegreen}{rgb}{0.33, 0.42, 0.18}
\definecolor{deepjunglegreen}{rgb}{0.0, 0.29, 0.29}
\definecolor{brickred}{rgb}{0.8, 0.25, 0.33}
\definecolor{deepjunglegreen}{rgb}{0.0, 0.29, 0.29}
\definecolor{darkpastelgreen}{rgb}{0.01, 0.75, 0.24}
\definecolor{green(pigment)}{rgb}{0.0, 0.65, 0.31}
\definecolor{junglegreen}{rgb}{0.16, 0.67, 0.53}
\definecolor{officegreen}{rgb}{0.0, 0.5, 0.0}
\definecolor{seagreen}{rgb}{0.18, 0.55, 0.34}
\definecolor{teal}{rgb}{0.0, 0.5, 0.5}
\definecolor{brightgreen}{rgb}{0.4, 1.0, 0.0}
\definecolor{electricgreen}{rgb}{0.0, 1.0, 0.0}
\definecolor{malachite}{rgb}{0.04, 0.85, 0.32}

\usepackage{accents}
\usepackage{colortbl}
\usepackage{verbatim}
\usepackage{float}
\usepackage{booktabs}
\usepackage{tikz}
\usetikzlibrary{arrows.meta,positioning}
\tikzset{
  negchg/.style={circle,draw,fill=cyan!8,inner sep=2.2pt},
  poschg/.style={circle,draw,fill=red!25,inner sep=2.2pt},
  vec/.style={-{Stealth[length=2.2mm]},thick},
  frc/.style={-{Stealth[length=2.2mm]},thick,red!70!black},
  vel/.style={-{Stealth[length=2.2mm]},thick,blue!80!black}
}

\makeatother

\providecommand\theoremname{Theorem}
\theoremstyle{plain}
\newtheorem{thm}{\protect\theoremname}[section]
\providecommand\remarkname{Remark}
\theoremstyle{remark}
\newtheorem{rem}[thm]{\protect\remarkname}
\providecommand\propositionname{Proposition}
\theoremstyle{plain}
\newtheorem{prop}[thm]{\protect\propositionname}
\providecommand\assumptionname{Assumption}
\newtheorem{assumption}[thm]{\protect\assumptionname}
\providecommand\definitionname{Definition}
\theoremstyle{definition}
\newtheorem{defn}[thm]{\protect\definitionname}
\providecommand\examplename{Example}
\newtheorem{example}[thm]{\protect\examplename}
\usepackage{babel}
\begin{document}

\global\long\def\ga{\alpha}%
\global\long\def\gb{\beta}%
\global\long\def\ggm{\gamma}%
\global\long\def\go{\omega}%
\global\long\def\gs{\sigma}%
\global\long\def\gd{\delta}%
\global\long\def\gD{\Delta}%
\global\long\def\vph{\phi}%
\global\long\def\gf{\phi}%
\global\long\def\gk{\kappa}%
\global\long\def\gl{\lambda}%
\global\long\def\gz{\zeta}%
\global\long\def\gh{\eta}%
\global\long\def\gy{\upsilon}%
\global\long\def\gth{\theta}%
\global\long\def\gO{\Omega}%
\global\long\def\gG{\Gamma}%

\global\long\def\eps{\varepsilon}%
\global\long\def\epss#1#2{\varepsilon^{#1}_{#2}}%
\global\long\def\ep#1{\eps_{#1}}%

\global\long\def\wh#1{\widehat{#1}}%
\global\long\def\hi{\hat{\imath}}%
\global\long\def\hj{\hat{\jmath}}%
\global\long\def\hk{\hat{k}}%
\global\long\def\ol#1{\overline{#1}}%
\global\long\def\ul#1{\underline{#1}}%

\global\long\def\spec#1{\textsf{#1}}%

\global\long\def\v#1{\boldsymbol{#1}}%

\global\long\def\ui{\wh{\boldsymbol{\imath}}}%
\global\long\def\uj{\wh{\boldsymbol{\jmath}}}%
\global\long\def\uk{\widehat{\boldsymbol{k}}}%

\global\long\def\uI{\widehat{\mathbf{I}}}%
\global\long\def\uJ{\widehat{\mathbf{J}}}%
\global\long\def\uK{\widehat{\mathbf{K}}}%

\global\long\def\mc#1{\mathcal{#1}}%
\global\long\def\bs#1{\boldsymbol{#1}}%
\global\long\def\vect#1{\mathbf{#1}}%
\global\long\def\bi#1{\textbf{\emph{#1}}}%

\global\long\def\uv#1{\widehat{\boldsymbol{#1}}}%
\global\long\def\cross{\times}%

\global\long\def\di{d}%
\global\long\def\dee#1{\mathop{d#1}}%

\global\long\def\ddt{\frac{\dee{}}{\dee t}}%
\global\long\def\dbyd#1{\frac{\dee{}}{\dee{#1}}}%
\global\long\def\dby#1#2{\frac{\partial#1}{\partial#2}}%
\global\long\def\dxdt#1{\frac{\dee{#1}}{\dee t}}%

\global\long\def\vct#1{\bs{#1}}%

\global\long\def\partialby#1#2{\frac{\partial#1}{\partial x^{#2}}}%
\newcommandx\parder[2][usedefault, addprefix=\global, 1=]{\frac{\partial#2}{\partial#1}}%
\global\long\def\supdot{^{\!\bs{\mathord{\cdot}}}}%

\global\long\def\fall{,\quad\text{for all}\quad}%

\global\long\def\reals{\mathbb{R}}%

\global\long\def\rthree{\reals^{3}}%
\global\long\def\rsix{\reals^{6}}%
\global\long\def\rn{\reals^{n}}%
\global\long\def\eucl{\mathbb{E}}%
\global\long\def\euthree{\eucl^{3}}%
\global\long\def\euln{\eucl^{n}}%

\global\long\def\prn{\reals^{n+}}%
\global\long\def\nrn{\reals^{n-}}%
\global\long\def\cprn{\overline{\reals}^{n+}}%
\global\long\def\cnrn{\overline{\reals}^{n-}}%
\global\long\def\rt#1{\reals^{#1}}%
\global\long\def\rtw{\reals^{12}}%

\global\long\def\les{\leqslant}%
\global\long\def\ges{\geqslant}%

\global\long\def\dX{\dee{\bp}}%
\global\long\def\dx{\dee x}%
\global\long\def\D{D}%

\global\long\def\from{\colon}%
\global\long\def\tto{\longrightarrow}%
\global\long\def\lmt{\longmapsto}%
\global\long\def\lhr{\lhook\joinrel\longrightarrow}%
\global\long\def\mto{\mapsto}%

\global\long\def\abs#1{\left|#1\right|}%

\global\long\def\isom{\cong}%

\global\long\def\comp{\circ}%

\global\long\def\cl#1{\overline{#1}}%

\global\long\def\fun{\varphi}%

\global\long\def\interior{\textrm{Int}\,}%
\global\long\def\inter#1{\kern0pt  #1^{\mathrm{o}}}%
\global\long\def\interior{\textrm{Int}\,}%
\global\long\def\inter#1{\kern0pt  #1^{\mathrm{o}}}%
\global\long\def\into{\mathrm{o}}%

\global\long\def\sign{\textrm{sign}\,}%
\global\long\def\sgn#1{(-1)^{#1}}%
\global\long\def\sgnp#1{(-1)^{\abs{#1}}}%

\global\long\def\du#1{#1^{*}}%

\global\long\def\tsum{{\textstyle \sum}}%
\global\long\def\lsum{{\textstyle \sum}}%

\global\long\def\dimension{\textrm{dim}\,}%

\global\long\def\esssup{\textrm{ess}\,\sup}%

\global\long\def\ess{\textrm{{ess}}}%

\global\long\def\kernel{\mathop{\textrm{\textup{Kernel}}}}%

\global\long\def\support{\mathop{\textrm{\textup{supp}}}}%

\global\long\def\image{\mathop{\textrm{\textup{Image}}}}%

\global\long\def\diver{\mathop{\textrm{\textup{div}}}}%

\global\long\def\spanv{\textrm{span}}%

\global\long\def\tr{\mathop{\textrm{\textup{tr}}}}%
\global\long\def\tran{\mathrm{tr}}%

\global\long\def\opt{\mathrm{opt}}%

\global\long\def\incl{\mathcal{I}}%
\global\long\def\iden{\imath}%
\global\long\def\idnt{\textrm{Id}}%
\global\long\def\rest{\rho}%
\global\long\def\extnd{e_{0}}%

\global\long\def\proj{\textrm{pr}}%

\global\long\def\L#1{L\bigl(#1\bigr)}%
\global\long\def\LS#1{L_{S}\bigl(#1\bigr)}%

\global\long\def\ino#1{\int_{#1}}%

\global\long\def\half{\frac{1}{2}}%
\global\long\def\shalf{{\scriptstyle \half}}%
\global\long\def\third{\frac{1}{3}}%

\global\long\def\empt{\varnothing}%

\global\long\def\innp#1#2{\left\langle #1,#2\right\rangle }%

\global\long\def\resto#1{|_{#1}}%
\global\long\def\compat#1#2{\left.#1\right|_{#2}}%

\global\long\def\paren#1{\left(#1\right)}%
\global\long\def\bigp#1{\bigl(#1\bigr)}%
\global\long\def\biggp#1{\biggl(#1\biggr)}%
\global\long\def\Bigp#1{\Bigl(#1\Bigr)}%

\global\long\def\braces#1{\left\{  #1\right\}  }%
\global\long\def\sqbr#1{\left[#1\right]}%
\global\long\def\anglep#1{\left\langle #1\right\rangle }%

\global\long\def\bigabs#1{\bigl|#1\bigr|}%
\global\long\def\dotp#1{#1^{\centerdot}}%
\global\long\def\pdot#1{#1^{\bs{\!\cdot}}}%

\global\long\def\eq{\sim}%
\global\long\def\quot{/\!\!\eq}%
\global\long\def\by{\!/\!}%

\global\long\def\stp{\text{{\small \ensuremath{\bigodot}}}}%
\global\long\def\tp{\text{{\small \ensuremath{\bigotimes}}}}%

\global\long\def\mi#1{#1}%
\global\long\def\mii{I}%
\global\long\def\mie#1#2{#1_{1}\cdots#1_{#2}}%

\global\long\def\smi#1{\boldsymbol{#1}}%
\global\long\def\asmi#1{#1}%
\global\long\def\ordr#1{\left\langle #1\right\rangle }%

\global\long\def\symm#1{\paren{#1}}%
\global\long\def\smtr{\mathcal{S}}%

\global\long\def\perm{p}%
\global\long\def\sperm{\mathcal{P}}%

\global\long\def\oneto{1,\dots,}%

\global\long\def\lisub#1#2#3{#1_{1}#2\dots#2#1_{#3}}%

\global\long\def\lisup#1#2#3{#1^{1}#2\dots#2#1^{#3}}%

\global\long\def\lisubb#1#2#3#4{#1_{#2}#3\dots#3#1_{#4}}%

\global\long\def\lisubbc#1#2#3#4{#1_{#2}#3\cdots#3#1_{#4}}%

\global\long\def\lisubbwout#1#2#3#4#5{#1_{#2}#3\dots#3\widehat{#1}_{#5}#3\dots#3#1_{#4}}%

\global\long\def\lisubc#1#2#3{#1_{1}#2\cdots#2#1_{#3}}%

\global\long\def\lisupc#1#2#3{#1^{1}#2\cdots#2#1^{#3}}%

\global\long\def\lisupp#1#2#3#4{#1^{#2}#3\dots#3#1^{#4}}%

\global\long\def\lisuppc#1#2#3#4{#1^{#2}#3\cdots#3#1^{#4}}%

\global\long\def\lisuppwout#1#2#3#4#5#6{#1^{#2}#3#4#3\wh{#1^{#6}}#3#4#3#1^{#5}}%

\global\long\def\lisubbwout#1#2#3#4#5#6{#1_{#2}#3#4#3\wh{#1}_{#6}#3#4#3#1_{#5}}%

\global\long\def\lisubwout#1#2#3#4{#1_{1}#2\dots#2\widehat{#1}_{#4}#2\dots#2#1_{#3}}%

\global\long\def\lisupwout#1#2#3#4{#1^{1}#2\dots#2\widehat{#1^{#4}}#2\dots#2#1^{#3}}%

\global\long\def\lisubwoutc#1#2#3#4{#1_{1}#2\cdots#2\widehat{#1}_{#4}#2\cdots#2#1_{#3}}%

\global\long\def\twp#1#2#3{\dee{#1}^{#2}\wedge\dee{#1}^{#3}}%

\global\long\def\thp#1#2#3#4{\dee{#1}^{#2}\wedge\dee{#1}^{#3}\wedge\dee{#1}^{#4}}%

\global\long\def\fop#1#2#3#4#5{\dee{#1}^{#2}\wedge\dee{#1}^{#3}\wedge\dee{#1}^{#4}\wedge\dee{#1}^{#5}}%

\global\long\def\idots#1{#1\dots#1}%
\global\long\def\icdots#1{#1\cdots#1}%

\global\long\def\norm#1{\|#1\|}%

\global\long\def\nonh{\heartsuit}%

\global\long\def\nhn#1{\norm{#1}^{\nonh}}%

\global\long\def\bigmid{\,\bigl|\,}%

\global\long\def\trps{^{{\scriptscriptstyle \textsf{T}}}}%

\global\long\def\testfuns{\mathcal{D}}%

\global\long\def\ntil#1{\tilde{#1}{}}%

\global\long\def\pis{y}%
\global\long\def\xo{\pis_{0}}%
\global\long\def\x{x}%

\global\long\def\pib{x}%
\global\long\def\bp{X}%
\global\long\def\ii{i}%
\global\long\def\ia{\alpha}%
\global\long\def\fp{y}%
\global\long\def\piv{v}%

\global\long\def\ib{i}%
\global\long\def\is{\alpha}%

\global\long\def\pbndo{\Gamma}%
\global\long\def\bndoo{\pbndo_{0}}%
 
\global\long\def\bndot{\pbndo_{t}}%
\global\long\def\intb{\inter{\body}}%
\global\long\def\bndb{\bdry\body}%

\global\long\def\cloo{\cl{\gO}}%

\global\long\def\nor{\nu}%
\global\long\def\Nor{\mathbf{N}}%

\global\long\def\dA{\dee A}%

\global\long\def\dV{\dee V}%

\global\long\def\eps{\varepsilon}%

\global\long\def\tv{v}%
\global\long\def\av{u}%

\global\long\def\svs{\mathcal{W}}%
\global\long\def\vs{\mathbf{V}}%
\global\long\def\avs{\mathbf{U}}%
\global\long\def\affsp{\mathcal{A}}%
\global\long\def\man{\mathcal{M}}%
\global\long\def\odman{\mathcal{N}}%
\global\long\def\subman{\mathcal{V}}%
\global\long\def\pt{p}%

\global\long\def\vbase{e}%
\global\long\def\sbase{\v e}%
\global\long\def\msbase{\mathfrak{e}}%
\global\long\def\vect{v}%
\global\long\def\dbase{\sbase}%

\global\long\def\chart{\varphi}%
\global\long\def\Chart{\Phi}%

\global\long\def\mind{\alpha}%
\global\long\def\vb{W}%
\global\long\def\vbp{\pi}%

\global\long\def\vbt{\mathcal{E}}%
\global\long\def\fib{\vs}%
\global\long\def\vbts{W}%
\global\long\def\avb{U}%
\global\long\def\vbp{\xi}%

\global\long\def\chart{\vph}%
\global\long\def\vbchart{\Phi}%

\global\long\def\jetb#1{J^{#1}}%
\global\long\def\jet#1{j^{1}(#1)}%
\global\long\def\tjet{\tilde{\jmath}}%

\global\long\def\Jet#1{J^{1}(#1)}%

\global\long\def\jetm{j}%

\global\long\def\coj{\mathfrak{d}}%

\global\long\def\alt{\mathfrak{A}}%

\global\long\def\pou{\eta}%

\global\long\def\ext{{\textstyle \bigwedge}}%
\global\long\def\forms{\Omega}%

\global\long\def\dotwedge{\dot{\mbox{\ensuremath{\wedge}}}}%

\global\long\def\vel{\theta}%

\global\long\def\Jac{\mathcal{J}}%

\global\long\def\contr{\mathbin{\raisebox{0.4pt}{\mbox{\ensuremath{\lrcorner}}}}}%
\global\long\def\fcor{\llcorner}%
\global\long\def\bcor{\lrcorner}%
\global\long\def\fcontr{\mathbin{\raisebox{0.4pt}{\mbox{\ensuremath{\llcorner}}}}}%

\global\long\def\lie{\mathcal{L}}%

\global\long\def\ssym#1#2{\ext^{#1}T^{*}#2}%

\global\long\def\sh{^{\sharp}}%

\global\long\def\nfo{\ext^{n}T^{*}\base}%
\global\long\def\dfs{\ext^{d}T^{*}\base}%
\global\long\def\dmfs{\ext^{d-1}T^{*}\base}%

\global\long\def\spc{\mathcal{S}}%
\global\long\def\sptm{\mathcal{E}}%
\global\long\def\evnt{e}%
\global\long\def\frame{\Psi}%

\global\long\def\timeman{\mathcal{T}}%
\global\long\def\zman{t}%
\global\long\def\dims{n}%
\global\long\def\m{\dims-1}%
\global\long\def\dimw{m}%

\global\long\def\wc{z}%

\global\long\def\fourv#1{\mbox{\ensuremath{\mathfrak{#1}}}}%

\global\long\def\body{\mathcal{B}}%
\global\long\def\man{\mathcal{M}}%
\global\long\def\var{\mathcal{V}}%
\global\long\def\base{\mathcal{X}}%
\global\long\def\fb{\mathcal{Y}}%
\global\long\def\srfc{\mathcal{Z}}%
\global\long\def\dimb{n}%
\global\long\def\dimf{m}%
\global\long\def\afb{\mathcal{Z}}%

\global\long\def\bdry{\partial}%

\global\long\def\gO{\varOmega}%

\global\long\def\reg{\gO}%
\global\long\def\bdrr{\bdry\reg}%

\global\long\def\bdom{\bdry\gO}%

\global\long\def\bndo{\partial\gO}%

\global\long\def\tpr{\vartheta}%

\global\long\def\mot{M}%
\global\long\def\vf{w}%
\global\long\def\const{h}%

\global\long\def\avf{u}%

\global\long\def\stn{\varepsilon}%
\global\long\def\djet{\chi}%

\global\long\def\jvf{\eps}%

\global\long\def\rig{r}%

\global\long\def\rigs{\mathcal{R}}%

\global\long\def\qrigs{\!/\!\rigs}%

\global\long\def\qd{\!/\,\!\kernel\diffop}%

\global\long\def\dis{\chi}%
\global\long\def\conf{\kappa}%
\global\long\def\invc{\hat{\conf}^{-1}}%
\global\long\def\dinvc{\hat{\conf}^{-1*}}%
\global\long\def\csp{\mathcal{Q}}%

\global\long\def\embds{\textrm{Emb}}%

\global\long\def\lc{A}%

\global\long\def\lv{\dot{A}}%
\global\long\def\alv{\dot{B}}%

\global\long\def\j{\mathop{\mathrm{j}}}%
\global\long\def\mapp{M}%
\global\long\def\J{J}%
\global\long\def\jex{\mathop{}\!\mathrm{j}}%

\global\long\def\fc{F}%
\global\long\def\load{f}%
\global\long\def\afc{g}%

\global\long\def\bfc{\mathbf{b}}%
\global\long\def\bfcc{b}%

\global\long\def\sfc{\mathbf{t}}%
\global\long\def\sfcc{t}%

\global\long\def\stm{\varsigma}%
\global\long\def\std{S}%
\global\long\def\tst{\sigma}%
\global\long\def\tstd{s}%
\global\long\def\st{\sigma}%
\global\long\def\vst{\varsigma}%
\global\long\def\vstd{S}%
\global\long\def\tstm{\sigma}%
\global\long\def\vstm{\varsigma}%

\global\long\def\stp{S_{P}}%
\global\long\def\slf{R}%

\global\long\def\crel{\Phi}%

\global\long\def\stmat{\tau}%

\global\long\def\gdiv{\bdry\textrm{iv\,}}%
\global\long\def\extjet{\mathfrak{d}}%

\global\long\def\smc#1{\mathfrak{#1}}%

\global\long\def\nhs{P}%
\global\long\def\nhsa{P}%
\global\long\def\nhsb{\underline{P}}%

\global\long\def\soc{Z}%

\global\long\def\sts{\varSigma}%
\global\long\def\spstd{\mathfrak{S}}%
\global\long\def\sptst{\mathfrak{T}}%
\global\long\def\spnhs{\mathcal{P}}%
\global\long\def\Ljj{\L{J^{1}(J^{k-1}\vb),\ext^{n}T^{*}\base}}%

\global\long\def\spsb{\text{{\Large \ensuremath{\Delta}}}}%

\global\long\def\ened{\mathfrak{w}}%
\global\long\def\energy{\mathfrak{W}}%

\global\long\def\ebdfc{T}%
\global\long\def\optimum{\st^{\textrm{opt}}}%
\global\long\def\scf{K}%

\global\long\def\grp{G}%
\global\long\def\gact{A}%
\global\long\def\gid{e}%
\global\long\def\gel{\ggm}%

\global\long\def\ael{\upsilon}%
\global\long\def\lal{\mathfrak{g}}%

\global\long\def\expr{\Pi}%

\global\long\def\aprop{Q}%

\global\long\def\flux{\omega}%
\global\long\def\aflux{\psi}%

\global\long\def\fform{\tau}%

\global\long\def\dimn{n}%

\global\long\def\sdim{{\dimn-1}}%

\global\long\def\fdens{\phi}%

\global\long\def\pform{s}%
\global\long\def\vform{\beta}%
\global\long\def\sform{\tau}%
\global\long\def\flow{\vf}%
\global\long\def\n{\m}%
\global\long\def\cmap{\mathfrak{t}}%
\global\long\def\vcmap{\varSigma}%

\global\long\def\mvec{\mathfrak{v}}%
\global\long\def\mveco#1{\mathfrak{#1}}%
\global\long\def\mv#1{\mathfrak{#1}}%
\global\long\def\smbase{\mathfrak{e}}%
\global\long\def\spx{\simp}%
\global\long\def\il{l}%
\global\long\def\awe{\frown}%

\global\long\def\hp{H}%
\global\long\def\ohp{h}%

\global\long\def\hps{G_{\dims-1}(T\spc)}%
\global\long\def\ohps{G^{\perp}_{\dims-1}(T\spc)}%

\global\long\def\hyper{\mathcal{S}}%

\global\long\def\hpsx{G_{\dims-1}(\tspc)}%
\global\long\def\ohpsx{G^{\perp}_{\dims-1}(\tspc)}%

\global\long\def\fbun{F}%

\global\long\def\flowm{\Phi}%

\global\long\def\tgb{T\spc}%
\global\long\def\ctgb{T^{*}\spc}%
\global\long\def\tspc{T_{\pis}\spc}%
\global\long\def\dspc{T^{*}_{\pis}\spc}%

\global\long\def\fflow{\fourv J}%
\global\long\def\fvform{\mathfrak{b}}%
\global\long\def\fsform{\mathfrak{t}}%
\global\long\def\fpform{\mathfrak{s}}%
\global\long\def\lfc{\mathfrak{F}}%

\global\long\def\maxw{\mathfrak{g}}%
\global\long\def\frdy{\mathfrak{f}}%
\global\long\def\ptnl{\gf}%
\global\long\def\pts{\Psi}%
\global\long\def\tptn{\Psi}%
\global\long\def\vptn{\mathfrak{a}}%
\global\long\def\mtst{\tstd_{M}}%
\global\long\def\mvst{\vstd_{M}}%

\global\long\def\sobp#1#2{W^{#1}_{#2}}%

\global\long\def\inner#1#2{\left\langle #1,#2\right\rangle }%

\global\long\def\fields{\sobp pk(\vb)}%

\global\long\def\bodyfields{\sobp p{k_{\partial}}(\vb)}%

\global\long\def\forces{\sobp pk(\vb)^{*}}%

\global\long\def\bfields{\sobp p{k_{\partial}}(\vb\resto{\bndo})}%

\global\long\def\loadp{(\sfc,\bfc)}%

\global\long\def\strains{\lp p(\jetb k(\vb))}%

\global\long\def\stresses{\lp{p'}(\jetb k(\vb)^{*})}%

\global\long\def\diffop{D}%

\global\long\def\strainm{E}%

\global\long\def\incomps{\vbts_{\yieldf}}%

\global\long\def\devs{L^{p'}(\eta^{*}_{1})}%

\global\long\def\incompsns{L^{p}(\eta_{1})}%

\global\long\def\testf{\mathcal{D}}%
\global\long\def\dists{\mathcal{D}'}%

\global\long\def\codiv{\boldsymbol{\partial}}%

\global\long\def\currof#1{\tilde{#1}}%

\global\long\def\chn{c}%
\global\long\def\chnsp{\mathbf{C}}%

\global\long\def\current{T}%
\global\long\def\curr{R}%

\global\long\def\curd{S}%
\global\long\def\curwd#1{\wh{#1}}%
\global\long\def\curnd#1{\wh{#1}}%

\global\long\def\contrf{{\scriptstyle \smallfrown}}%

\global\long\def\prodf{{\scriptstyle \smallsmile}}%

\global\long\def\form{\omega}%

\global\long\def\dens{\rho}%

\global\long\def\simp{s}%
\global\long\def\ssimp{\Delta}%
\global\long\def\cpx{K}%

\global\long\def\cell{C}%

\global\long\def\chain{B}%
\global\long\def\A{A}%
\global\long\def\B{B}%

\global\long\def\ach{A}%

\global\long\def\coch{X}%

\global\long\def\scale{s}%

\global\long\def\fnorm#1{\norm{#1}^{\flat}}%

\global\long\def\chains{\mathcal{A}}%

\global\long\def\ivs{\boldsymbol{U}}%

\global\long\def\mvs{\boldsymbol{V}}%

\global\long\def\cvs{\boldsymbol{W}}%

\global\long\def\ndual#1{#1'}%

\global\long\def\nd{'}%

\global\long\def\cee#1{C^{#1}}%

\global\long\def\lone{\{L^{1}\}}%

\global\long\def\linf{L^{\infty}}%

\global\long\def\lp#1{L^{#1}}%

\global\long\def\ofbdo{(\bndo)}%

\global\long\def\ofclo{(\cloo)}%

\global\long\def\vono{(\gO,\rthree)}%

\global\long\def\lomu{\{L^{1,\mu}\}}%
\global\long\def\limu{L^{\infty,\mu}}%
\global\long\def\limub{\limu(\body,\rthree)}%
\global\long\def\lomub{\lomu(\body,\rthree)}%

\global\long\def\vonbdo{(\bndo,\rthree)}%
\global\long\def\vonbdoo{(\bndoo,\rthree)}%
\global\long\def\vonbdot{(\bndot,\rthree)}%

\global\long\def\vonclo{(\cl{\gO},\rthree)}%

\global\long\def\strono{(\gO,\reals^{6})}%

\global\long\def\sob{\{W^{1}_{1}\}}%

\global\long\def\sobb{\sob(\gO,\rthree)}%

\global\long\def\lob{\lone(\gO,\rthree)}%

\global\long\def\lib{\linf(\gO,\reals^{12})}%

\global\long\def\ofO{(\gO)}%

\global\long\def\oneo{{1,\gO}}%
\global\long\def\onebdo{{1,\bndo}}%
\global\long\def\info{{\infty,\gO}}%

\global\long\def\infclo{{\infty,\cloo}}%

\global\long\def\infbdo{{\infty,\bndo}}%
\global\long\def\lobdry{\lone(\bdry\gO,\rthree)}%

\global\long\def\ld{LD}%

\global\long\def\ldo{\ld\ofO}%
\global\long\def\ldoo{\ldo_{0}}%

\global\long\def\trace{\gamma}%
\global\long\def\dtrace{\delta}%
\global\long\def\gtrace{\beta}%

\global\long\def\pr{\proj_{\rigs}}%

\global\long\def\pq{\proj}%

\global\long\def\qr{\,/\,\reals}%

\global\long\def\aro{S_{1}}%
\global\long\def\art{S_{2}}%

\global\long\def\mo{m_{1}}%
\global\long\def\mt{m_{2}}%

\global\long\def\ebdfc{T}%

\global\long\def\mini{\Omega}%
\global\long\def\optimum{s^{\mathrm{opt}}}%
\global\long\def\scf{K}%
\global\long\def\opsf{\st^{\mathrm{opt}}}%
\global\long\def\doptimum{s^{\opt,{\scriptscriptstyle D}}}%
\global\long\def\loptimum{s^{\opt,{\scriptscriptstyle \mathcal{M}}}}%

\global\long\def\fsubs{M}%

\global\long\def\yieldc{B}%

\global\long\def\yieldf{Y}%

\global\long\def\trpr{\pi_{P}}%

\global\long\def\devpr{\pi_{\devsp}}%

\global\long\def\prsp{P}%

\global\long\def\devsp{D}%

\global\long\def\ynorm#1{\|#1\|_{\yieldf}}%

\global\long\def\colls{\Psi}%

\global\long\def\aro{S_{1}}%
\global\long\def\art{S_{2}}%

\global\long\def\mo{m_{1}}%
\global\long\def\mt{m_{2}}%

\global\long\def\trps{^{\mathsf{T}}}%

\global\long\def\hb{^{\mathrm{hb}}}%

\global\long\def\yieldst{s_{Y}}%

\global\long\def\yieldc{B}%

\global\long\def\lcap{C}%

\global\long\def\yieldf{Y}%

\global\long\def\sphpr{\pi_{P}}%

\global\long\def\devpr{\pi_{\devsp}}%

\global\long\def\prsp{P}%

\global\long\def\devsp{D}%

\global\long\def\ynorm#1{\|#1\|_{\yieldf}}%

\global\long\def\colls{\Psi}%

\global\long\def\cone{Q}%
\global\long\def\fpr{\Pi}%
\global\long\def\fprd{\fpr_{\devsp}}%
\global\long\def\fprp{\fpr_{\prsp}}%
\global\long\def\find{I_{\devsp}}%
\global\long\def\finp{I_{\prsp}}%
\global\long\def\fnorm#1{\norm{#1}_{\devsp}}%

\global\long\def\rig{r}%
\global\long\def\rigs{\mathcal{R}}%
\global\long\def\qrigs{\!/\!\rigs}%
\global\long\def\anv{\omega}%
\global\long\def\I{I}%
\global\long\def\mone{M_{1}}%

\global\long\def\bd{BD}%

\global\long\def\po{\proj_{0}}%
\global\long\def\normp#1{\norm{#1}'_{\ld}}%

\global\long\def\ssx{S}%

\global\long\def\smap{s}%

\global\long\def\smat{\chi}%

\global\long\def\sx{e}%

\global\long\def\snode{P}%
\global\long\def\newmacroname{\{\}}%

\global\long\def\elem{e}%

\global\long\def\nel{L}%

\global\long\def\el{l}%

\global\long\def\gr{g}%
\global\long\def\ngr{G}%

\global\long\def\eldof{\alpha}%

\global\long\def\glbs{\psi}%

\global\long\def\ipln{\phi}%

\global\long\def\ndof{D}%

\global\long\def\dof{d}%

\global\long\def\nldof{N}%

\global\long\def\ldof{n}%

\global\long\def\lvf{\chi}%

\global\long\def\amat{A}%
\global\long\def\bmat{B}%

\global\long\def\subsp{\mathcal{M}}%
\global\long\def\zerofn{Z}%

\global\long\def\snomat{E}%

\global\long\def\femat{E}%

\global\long\def\tmat{T}%

\global\long\def\fvec{f}%

\global\long\def\snsp{\mathcal{S}}%

\global\long\def\slnsp{\Phi}%
\global\long\def\dslnsp{\Phi^{{\scriptscriptstyle D}}}%

\global\long\def\ro{r_{1}}%

\global\long\def\rtwo{r_{2}}%

\global\long\def\rth{r_{3}}%

\global\long\def\fmax{M}%

\global\long\def\dform{\psi}%

\global\long\def\srfc{\mathcal{S}}%

\global\long\def\semib{\mathrm{SB}}%

\global\long\def\tm#1{\overrightarrow{#1}}%
\global\long\def\tmm#1{\underrightarrow{\overrightarrow{#1}}}%

\global\long\def\itm#1{\overleftarrow{#1}}%
\global\long\def\itmm#1{\underleftarrow{\overleftarrow{#1}}}%

\global\long\def\ptrac{\mathcal{P}}%

\global\long\def\nh#1{\hat{#1}}%
\global\long\def\nj{\hat{\jmath}}%
\global\long\def\nJ{\hat{J}}%
\global\long\def\rin#1{\mathfrak{#1}}%
\global\long\def\npi{\hat{\pi}}%
\global\long\def\rp{\rin p}%
\global\long\def\rq{\rin q}%
\global\long\def\rr{\rin r}%

\global\long\def\xty{(\base,\fb)}%
\global\long\def\xts{(\base,\spc)}%
\global\long\def\r{r}%
\global\long\def\ntm{(\reals^{n},\reals^{m})}%

\global\long\def\tproj{\frame_{\timeman}}%
\global\long\def\sproj{\frame_{\spc}}%

\global\long\def\cons{c}%
\global\long\def\optm{\go}%
\global\long\def\flxs{\mathcal{W}}%
\global\long\def\cost{Q}%

\global\long\def\mtn{e}%
\global\long\def\sppp{\lambda}%

\global\long\def\mtsp{\mathscr{E}}%

\global\long\def\disp{g}%
\global\long\def\diffs{G}%

\global\long\def\bv{BV}%

\global\long\def\Charge{Q}%

\global\long\def\pole{q}%

\global\long\def\pdens{\rho}%

\global\long\def\ms#1{\mathfrak{#1}}%

\global\long\def\Hfl{\Phi}%

\global\long\def\cud{\v j}%
\global\long\def\mag{\v M}%
\global\long\def\vp{\v A}%
\global\long\def\magf{\v B}%
\global\long\def\magi{\v H}%

\global\long\def\pfun{\mathscr{P}}%

\title[Magnetostatics as a Stress Theory]{Magnetostatics as a Stress Theory in $\rthree$: Free and Bound Currents,
Gauge Invariance, and Mechanical Interactions}
\author{Vladimir Gol'dshtein, Wolfgang H.~Müller, and Reuven Segev}
\date{\today}
\begin{abstract}
We formulate magnetostatics in $\rthree$ as a theory of stresses
in which the virtual velocity field of continuum mechanics is replaced
by the vector potential $\v A$. An electric current system is a collection
of linear functionals on vector potentials, one for each subbody,
the values of which are interpreted as minus the potential energy.
Each functional is represented by a volume current density and by
a surface current density on the boundary of the subbody. Cauchy's
theorem gives a tensor field $\v{\Sigma}$, the analog of the traction
stress, with the surface current $\v{\Sigma}\v{\nor}$ (the action
of $\v{\Sigma}$, viewed as a linear mapping, on $\v{\nor}$), where
$\v{\nor}$ is the outward unit normal to the boundary. We do not
assume that the energy is gauge invariant. We only assume weak gauge
invariance. That is, a gauge transformation $\v A\mapsto\v A+\nabla\gl$,
for a gauge function $\gl$, changes the energy only through the values
of $\nabla\gl$. This is the analog of the symmetry of the stress,
and we show that it holds if and only if $\v{\Sigma}$ is skew-symmetric,
so that it is represented by a vector field $\v M$. The energy then
takes the form $-\int(\v j_{f}\cdot\v A+\v M\cdot\v B)\dV$, which
splits every current system uniquely into free current $\v j_{f}$
and a distribution of magnetic dipoles with density $\v M$, carrying
the bulk and boundary bound currents $\nabla\cross\v M$ and $\v M\cross\v{\nor}$.
Gauge invariance of the energy is the analog of the balance of forces,
and it singles out the current systems without free current. For currents
embedded in matter, the forces exerted by the magnetic field are derived.
They consist of the Lorentz force on the total current, free and bound,
and of a gauge-dependent term that appears only where free current
crosses the boundary of the subbody. Force functionals also include
an asymmetric active stress $(\v M\cdot\v B)\v I-\v M\otimes\v B$
whose skew-symmetric part is the couple density $\v M\cross\v B$.
We show that the gauge-dependent term is the rate at which the moving
charge carriers transport potential momentum out of the subbody, so
that the force on the matter itself is the gauge-invariant part.
\end{abstract}

\maketitle
\tableofcontents{}

\section{Introduction}

Standard treatments of the magnetostatics of ponderable bodies (e.g.,
{[}Jac99, Zan13{]}) introduce a magnetization field, split the current
into free and bound parts, and then postulate the force and the couple
acting on magnetized matter. In particular, the fields are the magnetic
flux density $\v B$, the magnetic field $\v H$, the magnetization
$\v M$, and the free current density $\v j_{f}$. The field equations
are
\[
\nabla\cdot\v B=0,\qquad\nabla\cross\v H=\v j_{f},
\]
and they are completed by constitutive relations. The relation
\[
\v H=\frac{1}{\mu_{0}}\v B-\v M,
\]
where $\mu_{0}$ is the permeability of vacuum, may be viewed either
as the definition of $\v H$ or as the aether relation of the vacuum
{[}TT60, Sect.~279{]}, modified by the magnetization of the matter.
The response of the material is described by a constitutive relation
for $\v M$, for example, $\v M=\chi_{m}\v H$ for a linear material
with magnetic susceptibility $\chi_{m}$. The general problem is to
find, for a given free current and a given material, fields $\v B$,
$\v H$ and $\v M$ that satisfy the field equations and the constitutive
relations, together with the jump conditions at interfaces and the
decay of the fields at infinity. The force and the couple on the matter
are then postulated, and, in particular, the Lorentz force is postulated
as the force exerted by the field on the currents. Here, without using
these equations or any constitutive relations (such as the aether
relations), we adopt a different approach, based on the stress theory
of geometric continuum mechanics.

In a companion paper {[}GMS26a{]}, we applied this approach for electrostatics.
We regarded a charge distribution in a region $\gO\subset\rthree$
as a collection of linear functionals on potential fields, one for
each subbody of $\gO$. We showed that electrostatics is then the
simplest example of the theory of stresses, where the virtual velocity
is replaced by a scalar potential. Cauchy's theorem gives the split
into free charge and dipoles, and gauge invariance of the energy plays
the role of the balance of forces. Here we carry the construction
over to magnetostatics, where the generalized velocity is the vector
potential $\v A$. A new feature appears, and it is one of the main
points of the paper.

In electrostatics, every charge system is represented by a free charge
density and a dipole density. In magnetostatics, when Cauchy's theorem
is applied to the electric current system, it gives by analogy a tensor
field $\v{\Sigma}$. A general tensor has nine components, while the
magnetization field that will represent it has only three. The extra
components act on the second derivatives of the gauge function $\gl$
when $\v A$ is replaced by $\v A+\nabla\gl$ (for the gauge freedom
of the vector potential, see {[}PP62, p.~241{]}). We do not assume
that the energy is gauge invariant, since, as in electrostatics, a
body carrying free current does not have a gauge-invariant energy.
We assume only that a gauge transformation changes the energy through
the values of $\nabla\gl$ and not through its derivatives, i.e.,
not through the second derivatives of $\gl$. We call this property
weak gauge invariance (Assumption~\ref{ass:weak}). We show that
it is equivalent to the skew-symmetry of $\v{\Sigma}$, so that $\v{\Sigma}$
is represented by a vector field $\v M$, and that it is the exact
analog of the symmetry of the stress in continuum mechanics. The gauge
fields $\nabla\gl$, whose gradients are symmetric, assume the role
of the rigid velocity fields, whose gradients are skew-symmetric.

The forces and couples that the magnetic field exerts on magnetized
matter have a long history in continuum mechanics. Brown {[}Bro66{]}
developed the theory of magnetoelastic interactions, and comprehensive
treatments can be found in {[}Mau88, EM90, Kov00, DO14{]}. In these
works the couple $\v M\cross\v B$ exerted by the field makes the
stress in the material nonsymmetric. We do not aim at new field equations
or constitutive relations. What we offer is of a different nature:
(i) we treat the current system itself as a force system in the sense
of stress theory, so that the split into free current and magnetization
follows from Cauchy's theorem and weak gauge invariance instead of
being postulated; (ii) we show that weak gauge invariance is the analog
of the symmetry of the stress, and that gauge invariance of the energy
is the analog of the balance of forces, a way of classifying current
systems rather than a postulate; and (iii) the forces on embedded
currents, including the asymmetric stress and the couple $\v M\cross\v B$,
follow from the energy by a single differentiation, and we identify
the gauge-dependent part of the force as a transport of potential
momentum.

In the analogy with continuum mechanics:
\begin{itemize}
\item the electric current system plays the role of a force system, with
volume current as body force and surface current as surface traction;
\item a tensor field $\v{\Sigma}$ plays the role of the stress, and a vector
field $\v r$ plays the role of the unbalanced body force;
\item minus the potential energy plays the role of the virtual power;
\item gauge transformations $\v A\mapsto\v A+\nabla\gl$ play the role of
rigid motions.
\end{itemize}
For a weakly gauge-invariant current system (see Section~\ref{sec:currents}),
the energy of a subbody $\gO$ is 
\begin{equation}
U_{\gO}(\v A)=-\int_{\gO}\big(\v j_{f}\cdot\v A+\v M\cdot\v B\big)\dV,\qquad\v B:=\nabla\cross\v A,\label{eq:U-intro}
\end{equation}
the analog of the representation of a force system by a variational
stress {[}Seg02{]}, {[}Seg23, Ch.~10{]}. The pair $(\v j_{f},\v M)$
represents a unique decomposition of the electric current system into
free current and a distribution of magnetic dipoles. Requiring the
energy of every subbody to be gauge invariant is the analog of requiring
the forces on every subbody to add up to zero. It singles out the
current systems with $\v j_{f}=\v 0$, those consisting of closed
current loops at the microscopic level, for which $U_{\gO}(\v A)=-\int_{\gO}\v M\cdot\v B\dV$.

The mechanical forces show a second difference from electrostatics.
There, the forces on embedded charges\textemdash charges that follow
the motion of matter\textemdash are gauge invariant for every charge
system. Here, the forces on embedded currents\textemdash currents
embedded in matter and convected with its motion\textemdash consist
of the Lorentz force on the total current, which is gauge invariant,
and of a term that depends on the gauge and that appears only where
free current crosses the boundary of the subbody. This additional
term reflects the fact that the energy of an open piece of a circuit
is not gauge invariant. We show that this term is the rate at which
the moving charge carriers transport potential momentum out of the
subbody. It is not a force on the matter, and the force on the matter
is the Lorentz force on the total current, together with the force
on the bound surface current.

We start with a single current loop and a single current element in
Section~\ref{sec:dipole}. Section~\ref{sec:stress} recalls what
we need from the theory of stresses, including the notion of a variational
stress, its distinction from the traction stress, and the role of
rigid rotations. In Section~\ref{sec:currents} we introduce current
systems, weak gauge invariance, and the split into free and bound
currents. Section~\ref{sec:gauge} characterizes the current systems
without free current by gauge invariance, discusses the relation to
the conservation of charge, and distinguishes three levels of gauge
invariance. In Section~\ref{sec:mech} we derive the mechanical forces,
and Section~\ref{sec:examples} gives examples. We close with some
remarks in Section~\ref{sec:concl}.

Throughout, a comma denotes partial differentiation, we use the summation
convention, and we regard a second-order tensor $\v T$ as a linear
mapping of vectors, so that $\v T\v v$ denotes the vector with components
$(\v T\v v)_{i}=T_{ij}v_{j}$. The transpose is denoted by $\v T\trps$,
the tensor product of vectors by $(\v a\otimes\v b)_{ij}=a_{i}b_{j}$,
and $\v T:\v S=T_{ij}S_{ij}$. In addition, $\eps_{ijk}$ is the permutation
symbol, and the magnetic field is defined by $\v B:=\nabla\cross\v A$,
so that $\nabla\cdot\v B=0$. We will often use the identity 
\begin{equation}
\eps_{ijk}\eps_{ilm}=\gd_{jl}\gd_{km}-\gd_{jm}\gd_{kl},\label{eq:epsdelta}
\end{equation}
together with the invariance of $\eps_{ijk}$ under cyclic permutations
of its indices. We denote by $\v M$ the field that plays the role
of the magnetization, and the field $\v H$ appears only in Section~\ref{sec:concl}.
The gauge function is denoted by $\gl$, while $\dis$ denotes a virtual
motion.

\section{Motivation: a single magnetic dipole}

\label{sec:dipole}

\subsection{The naive point of view}

\textbf{The setting.} We work in the setting of magnetostatics. All
currents are steady. The current in the loop we consider, and the
currents in the external circuits that produce the magnetic field
$\v B$, are kept at fixed values by the sources (batteries) that
drive them, and the virtual motions are slow enough for the system
to remain magnetostatic at every instant. The field $\v B$ is the
external field. We do not consider the forces that the field of the
loop exerts on the loop itself.

In the naive picture, a magnetic dipole is a small planar loop $C$
carrying a current $I$ (Figure~\ref{fig:loop}). This is Ampère's
model of a magnetic dipole. Let $\v y$ be a point of the loop's plane
inside the loop, write the points of the loop as $\v y+\v{\xi}$,
and let $\v a$ be the vector area of the loop, oriented with the
current by the right-hand rule. We let the loop shrink and $I$ grow
so that the magnetic moment 
\begin{equation}
\v m:=I\,\v a\label{eq:m}
\end{equation}
remains finite. We use the identity 
\begin{equation}
\oint_{C}\xi_{l}\,\dee{\xi_{i}}=\eps_{lik}a_{k},\label{eq:loopid}
\end{equation}
which says that the projections of the loop on the coordinate planes
enclose the components of $\v a$. To prove it, let $S$ be the plane
region enclosed by $C$, and $\v{\nor}$ its unit normal, oriented
with $C$ by the right-hand rule, so that $\v a=\int_{S}\v{\nor}\dA$
is the area of $S$ times $\v{\nor}$. Since $\v e_{i}$ is constant,
$\nabla\cross(\xi_{l}\v e_{i})=\nabla\xi_{l}\cross\v e_{i}=\v e_{l}\cross\v e_{i}$,
and Stokes' theorem gives 
\[
\oint_{C}\xi_{l}\,\dee{\xi_{i}}=\oint_{C}(\xi_{l}\v e_{i})\cdot\dee{\v{\xi}}=\int_{S}(\v e_{l}\cross\v e_{i})\cdot\v{\nor}\dA=\eps_{lik}a_{k}.
\]
Since $\oint_{C}\dee{\v{\xi}}=0$, the identity does not depend on
the point from which $\v{\xi}$ is measured. For $l=1$, $i=2$, it
gives $\oint_{C}\xi_{1}\,\dee{\xi_{2}}=a_{3}$, which, by Green's
theorem, is the area enclosed by the projection of $C$ on the $\xi_{1}\xi_{2}$-plane.\footnote{The loop need not be planar. For an arbitrary closed loop $C$, define
$\v a:=\int_{S}\v{\nor}\dA$, where $S$ is any surface spanning $C$,
with $\v{\nor}$ oriented with $C$ by the right-hand rule. The vector
$\v a$ does not depend on the choice of $S$, since $\int\v{\nor}\dA=0$
over every closed surface. In general, it equals $\half\oint_{C}\v{\xi}\cross\dee{\v{\xi}}$.
With this definition, the proof above applies word for word, and (\ref{eq:loopid})
holds for every closed loop.}

\begin{figure}[ht]
\begin{tikzpicture}[scale=1.1]
\draw[thick] (0,0) ellipse (1.6 and 0.6);
\draw[vec] (0.15,-0.6) -- (0.25,-0.6);
\node at (1.0,-0.85) {$I$};
\fill (0,0) circle (1.2pt) node[below left] {$\v y$};
\draw[vec] (0,0) -- (0,1.4) node[right] {$\v m=I\v a$};
\draw[vec] (0,0) -- (1.13,0.42) node[midway,below right] {$\v{\xi}$};
\end{tikzpicture} \caption{A small current loop.}
\label{fig:loop}
\end{figure}

\textbf{The magnetic flux.} The magnetic flux through the loop due
to the external field $\v B$ is $\Phi=\int_{S}\v B\cdot\v{\nor}\dA$.
For a small loop, $\Phi\cong\v B(\v y)\cdot\v a$. We will, however,
write the flux in terms of the vector potential. By Stokes' theorem,
$\int_{S}(\nabla\cross\v A)\cdot\v{\nor}\dA=\oint_{C}\v A\cdot\dee{\v{\xi}}$,
so that 
\begin{equation}
\Phi=\oint_{C}\v A\cdot\dee{\v{\xi}}\cong\oint_{C}\big(A_{i}(\v y)+A_{i,l}(\v y)\,\xi_{l}\big)\dee{\xi_{i}}=\eps_{lik}a_{k}A_{i,l}(\v y)=\v a\cdot\v B(\v y),\label{eq:flux}
\end{equation}
where we used $\oint_{C}\dee{\xi_{i}}=0$, (\ref{eq:loopid}), and
$B_{k}=\eps_{kli}A_{i,l}$. There are two reasons for using $\v A$
rather than $\v B$. First, the vector potential is the generalized
velocity of the theory, and we want the energy as a functional of
$\v A$. Second, $\oint_{C}\v A\cdot\dee{\v{\xi}}$ is the sum of
contributions $\v A\cdot\dee{\v{\xi}}$ of the individual current
elements of the loop, and this form makes sense also for a current
element that is not part of a closed loop, for which there is no surface
$S$ and no flux of $\v B$. We will need such elements below.

\textbf{Virtual power for an embedded loop.} Now suppose that the
loop is embedded in a material body in virtual motion with velocity
field $\v w$, so that the wire moves with the material. The Lorentz
force on the current element $I\dee{\v{\xi}}$ is $I\dee{\v{\xi}}\cross\v B$,
and the virtual power of the forces exerted by the magnetic field
is 
\begin{equation}
W_{m}=I\oint_{C}(\dee{\v{\xi}}\cross\v B)\cdot\v w=I\oint_{C}\eps_{kij}\,\dee{\xi_{i}}\,(B_{j}w_{k})(\v y+\v{\xi}).\label{eq:Wloop}
\end{equation}
We expand $B_{j}w_{k}$ about $\v y$, 
\[
(B_{j}w_{k})(\v y+\v{\xi})=(B_{j}w_{k})(\v y)+(B_{j}w_{k})_{,l}(\v y)\,\xi_{l}+O(|\v{\xi}|^{2}).
\]
After substituting into the integral, the first term contributes nothing,
since $\oint_{C}\dee{\xi_{i}}=0$. The second gives, by (\ref{eq:loopid}),
\[
I\eps_{kij}(B_{j}w_{k})_{,l}\oint_{C}\xi_{l}\,\dee{\xi_{i}}=I\eps_{kij}\eps_{lin}a_{n}(B_{j}w_{k})_{,l}=m_{n}\eps_{kij}\eps_{lin}(B_{j}w_{k})_{,l}.
\]
The remainder is bounded by a constant times $I\ell^{3}$, where $\ell$
is the size of the loop. Since $I\ell^{2}$ is of the order of $|\v m|$,
which is kept fixed, the remainder vanishes in the limit. Hence, by
(\ref{eq:epsdelta}), in the limit, 
\begin{equation}
\begin{split}W_{m} & =m_{n}(B_{n}w_{k})_{,k}-m_{n}(B_{l}w_{n})_{,l}\\
 & =m_{n}B_{n,k}w_{k}+(\v m\cdot\v B)\,w_{k,k}-(\nabla\cdot\v B)\,\v m\cdot\v w-m_{n}B_{l}w_{n,l}.
\end{split}
\end{equation}
Since $\nabla\cdot\v B=0$, the third term drops out, and, renaming
the indices, 
\begin{equation}
\begin{split}W_{m} & =m_{i}B_{i,k}w_{k}+(\v m\cdot\v B)\,w_{k,k}-m_{i}B_{j}w_{i,j}\\
 & =\v f_{m}\cdot\v w+\big((\v m\cdot\v B)\v I-\v m\otimes\v B\big):\nabla\v w,
\end{split}
\label{eq:Wm}
\end{equation}
all fields evaluated at $\v y$, where, for a tensor $\v T$, $\v T:\nabla\v w=T_{ij}w_{i,j}$,
and 
\begin{equation}
\v f_{m}=(\nabla\v B)\trps\v m,\qquad(f_{m})_{k}=m_{i}B_{i,k}.\label{eq:fm}
\end{equation}
For a uniform velocity field only the first term survives, and it
gives the power of the total force $\v f_{m}$, the familiar $\nabla(\v m\cdot\v B)$
at a fixed moment. For a rigid rotation $\v w=\v{\go}\cross\v x$,
we have $w_{i}=\eps_{ikl}\go_{k}x_{l}$, so that $w_{i,j}=\eps_{ikj}\go_{k}$
and $w_{k,k}=0$. Since $\eps_{ikj}=-\eps_{kij}$, 
\begin{equation}
-m_{i}B_{j}w_{i,j}=\go_{k}\eps_{kij}m_{i}B_{j}=\v{\go}\cdot(\v m\cross\v B),\label{eq:couple-m}
\end{equation}
so the skew-symmetric part of the tensor $(\v m\cdot\v B)\v I-\v m\otimes\v B$
represents the couple $\v m\cross\v B$. As for the electric dipole
{[}GMS26a, Sect.~2{]}, the tensor acts on the velocity gradient to
produce power, so it is a stress concentrated at $\v y$, and it is
not symmetric.

\textbf{The potential energy.} We will show in Section~\ref{sec:dipole-functional}
that, for every virtual motion of the embedded loop, the virtual power
(\ref{eq:Wm}) of the Lorentz forces is the rate of increase of $I\Phi$,
where $\Phi$ is the flux through the moving and deforming loop and
$I$ is held fixed. In other words, at a fixed current, the magnetic
forces on the loop are derived from the potential energy 
\begin{equation}
U_{m}=-I\Phi=-I\oint_{C}\v A\cdot\dee{\v{\xi}}=-\v m\cdot\v B(\v y),\label{eq:Um}
\end{equation}
in the sense that their virtual power is $-\dee{U_{m}}/\dee t$. This
is what we assume about magnetostatics, and it is all we need. The
mechanical forces on fixed steady currents are derived from $U_{m}$.\footnote{It is worth being specific about what $U_{m}$ is not. It is not the
total energy of the system. When the loop moves at a fixed current
in the field of other fixed currents, the flux through each circuit
changes, and the sources must work against the induced electromotive
forces to keep the currents fixed. Keeping the external currents fixed
is the same as keeping the vector potential $\v A$ fixed in space,
as we do throughout the paper, since, up to a gauge transformation,
$\v A$ is determined by the external currents. One may think of $\v A$
as held fixed by a control system. Here, the control system consists
of the sources of the external circuits, while the source of the loop
keeps $I$ fixed, and the energy they supply is the price of this
control. In a small displacement, the sources supply the energy $2I\gd\Phi$,
of which the source of the loop supplies $I\gd\Phi$ and the sources
of the external circuits supply $I\gd\Phi$. Half of it, $I\gd\Phi$,
increases the energy of the magnetic field, and the other half is
the mechanical work $I\gd\Phi$ done by the magnetic forces {[}Jac99,
Sects.~5.7 and 5.16{]}. Thus, the field energy increases by $I\gd\Phi$,
while the mechanical work is $-\gd U_{m}=I\gd\Phi$. For a permanent
magnetic moment, for which no source is involved, $U_{m}=-\v m\cdot\v B$
is the usual potential energy of a dipole in an external field {[}Jac99,
Sect.~5.7{]}. It is in this sense, as a potential for the mechanical
forces at fixed currents, that we use the term potential energy.}

\subsection{Current elements and dipoles as linear functionals}

\label{sec:dipole-functional}

With (\ref{eq:Um}) in mind, we drop the limit of a shrinking loop
and define a magnetic dipole at $\v y$ as the linear functional on
$C^{1}$ vector potentials 
\begin{equation}
I_{m}(\v A):=\v m\cdot\v B(\v y)=\v m\cdot(\nabla\cross\v A)(\v y),\label{eq:Im}
\end{equation}
which we interpret as minus the potential energy of the dipole in
the field. Compare this with a current element $I\dee{\v l}$ at $\v y$,
by which we mean, using the term loosely, the functional 
\begin{equation}
I_{e}(\v A):=I\dee{\v l}\cdot\v A(\v y),\label{eq:Ie}
\end{equation}
whose energy $-I\dee{\v l}\cdot\v A(\v y)$ is the contribution of
the element to (\ref{eq:Um}). Two properties of the dipole functional,
which the current element does not share, will be central in what
follows.
\begin{enumerate}
\item[(a)] A dipole is gauge invariant. For any smooth function $\gl$, $I_{m}(\v A+\nabla\gl)=I_{m}(\v A)$,
since $\nabla\cross\nabla\gl=0$. For a current element, $I_{e}(\v A+\nabla\gl)=I_{e}(\v A)+I\dee{\v l}\cdot\nabla\gl(\v y)$,
and the change depends on $\nabla\gl$ only through its value at $\v y$.
\item[(b)] The energy of the dipole depends on the potential only through the
magnetic field: $I_{m}(\v A)=\v m\cdot\v B(\v y)$.
\end{enumerate}
\textbf{The embedded dipole.} Let the dipole be embedded in a body
undergoing a virtual motion $\dis\colon\reals\times\gO\to\rthree$.
We write $\dis_{t}:=\dis(t,\cdot)$ for the configuration at time
$t$, $\v F_{t}=\nabla\dis_{t}$ for its deformation gradient, with
components $F_{t\,ij}=\dis_{t\,i,j}$, and $J_{t}=\det\v F_{t}$ for
its Jacobian. We assume that $\dis_{0}=\mathrm{Id}$, and we let the
corresponding virtual velocity field be $\v w=\bdry_{t}\dis_{t}|_{t=0}$.
Then 
\begin{equation}
\v F_{0}=\v I,\qquad\bdry_{t}\v F_{t}|_{t=0}=\nabla\v w,\qquad\bdry_{t}J_{t}|_{t=0}=w_{k,k}.\label{eq:Fdot}
\end{equation}
The first is $\dis_{0}=\mathrm{Id}$, the second follows by interchanging
the derivatives with respect to $t$ and $\v x$, and the third is
the derivative of the determinant at the identity, $\bdry_{t}\det\v F_{t}|_{t=0}=\tr\,\bdry_{t}\v F_{t}|_{t=0}$.

The current $I$ in the wire is not affected by the motion, and the
loop spans a material area element, which is deformed by the motion.
By Nanson's formula, the area element is carried as $\v a_{t}=J_{t}\v F^{-\mathsf{T}}_{t}\v a$,
so that 
\begin{equation}
\v m_{t}=J_{t}\v F^{-\mathsf{T}}_{t}(\v y)\,\v m.\label{eq:mt}
\end{equation}
Thus, a magnetic moment is carried as an area element, whereas the
electric dipole of {[}GMS26a{]} is carried as a line element. To differentiate
(\ref{eq:mt}), we differentiate $\v F_{t}\v F^{-1}_{t}=\v I$, which
gives $\bdry_{t}(\v F^{-1}_{t})=-\v F^{-1}_{t}(\bdry_{t}\v F_{t})\v F^{-1}_{t}$,
and at $t=0$, by (\ref{eq:Fdot}), $\bdry_{t}(\v F^{-1}_{t})|_{t=0}=-\nabla\v w$.
Transposing, and using the product rule with $\bdry_{t}J_{t}|_{t=0}=w_{k,k}$,
\begin{equation}
\begin{split}\bdry_{t}(J_{t}\v F^{-\mathsf{T}}_{t})|_{t=0} & =w_{k,k}\v I-(\nabla\v w)\trps,\\
\bdry_{t}(J_{t}\v F^{-\mathsf{T}}_{t})_{ij}|_{t=0} & =w_{k,k}\gd_{ij}-w_{j,i}.
\end{split}
\label{eq:JFTdot}
\end{equation}
The energy at time $t$ is $U_{t}=-\v m_{t}\cdot\v B(\dis_{t}(\v y))=-(J_{t}\v F^{-\mathsf{T}}_{t})_{ij}m_{j}B_{i}(\dis_{t}(\v y))$.
By the product rule, (\ref{eq:JFTdot}), and the chain rule $\bdry_{t}B_{i}(\dis_{t}(\v y))|_{t=0}=B_{i,k}(\v y)w_{k}(\v y)$,
\[
\frac{\dee{}}{\dee t}U_{t}\Big|_{t=0}=-(w_{k,k}\gd_{ij}-w_{j,i})m_{j}B_{i}-m_{i}B_{i,k}w_{k}.
\]
We define the force on the dipole as the functional 
\begin{equation}
F_{m}(\v w):=-\frac{\dee{}}{\dee t}U_{t}\Big|_{t=0}=m_{i}B_{i,k}w_{k}+(\v m\cdot\v B)\,w_{k,k}-m_{i}B_{j}w_{i,j},\label{eq:Fm}
\end{equation}
where, in the last term, we renamed $i\leftrightarrow j$. All fields
are evaluated at $\v y$.

It follows that $F_{m}(\v w)=W_{m}$. That is, the virtual power of
the Lorentz forces (\ref{eq:Wm}) is $-\dee{U_{t}}/\dee t$, as stated
above. The two expressions are the same, but they were obtained differently,
and there is one difference worth noting. The computation through
the Lorentz forces produced the additional term $-(\nabla\cdot\v B)\,\v m\cdot\v w$,
which we dropped because $\nabla\cdot\v B=0$. The computation through
the energy never produces it. Hence, the two agree because $\v B=\nabla\cross\v A$
is divergence free. For a field with nonvanishing divergence, which
cannot be the curl of a potential, they would differ.
\begin{rem}
\textbf{Ampère dipoles and electron spins.} The present formulation
describes magnetic dipoles as small current loops embedded in the
material. For such Ampère dipoles, the transformation law (\ref{eq:mt}),
and its continuum counterpart $\v M_{t}=\v F^{-\mathsf{T}}_{t}\v M$
in (\ref{eq:convect}), follow from the deformation of the oriented
area enclosed by the loop, provided that the current in the loop remains
constant. This assumption is less natural for intrinsic magnetic moments,
such as electron spins. An electron spin is not a classical current
loop whose enclosed area changes under deformation. Its magnitude
is fixed by quantum mechanics, while its orientation and its coupling
to the material depend on microscopic interactions, including spin\textendash orbit
coupling and magnetic anisotropy. Thus, the transformation law is
justified directly for embedded Ampère dipoles, whereas its application
to spin magnetization requires additional physical or constitutive
assumptions. The distinction is relevant when the magnetic stresses
and couples derived from the energy are interpreted for magnetized
materials.
\end{rem}

\textbf{The embedded current element.} A current element $I\dee{\v l}$
in a wire attached to the body is carried as a material line element,
$\dee{\v l_{t}}=\v F_{t}\dee{\v l}$, with the same current. Its energy
is 
\begin{equation}
U_{t}=-I\dis_{t\,i,j}(\v y)\,\dee{l_{j}}\,A_{i}(\dis_{t}(\v y)),\label{eq:Ue}
\end{equation}
and, by (\ref{eq:Fdot}) and the chain rule, 
\begin{equation}
F_{e}(\v w):=-\frac{\dee{}}{\dee t}U_{t}\Big|_{t=0}=I\big(A_{i}w_{i,j}\dee{l_{j}}+\dee{l_{i}}A_{i,k}w_{k}\big).
\end{equation}
To see the Lorentz force in this expression, we compute $\dee{\v l}\cross\v B$.
With $B_{j}=\eps_{jpq}A_{q,p}$ and, by cyclic permutation and (\ref{eq:epsdelta}),
$\eps_{kij}\eps_{jpq}=\eps_{jki}\eps_{jpq}=\gd_{kp}\gd_{iq}-\gd_{kq}\gd_{ip}$,
\begin{equation}
(\dee{\v l}\cross\v B)_{k}=\eps_{kij}\dee{l_{i}}B_{j}=\eps_{kij}\eps_{jpq}\dee{l_{i}}A_{q,p}=\dee{l_{i}}(A_{i,k}-A_{k,i}).\label{eq:dlxB}
\end{equation}
On the other hand, $\dee{l_{j}}(A_{i}w_{i})_{,j}=\dee{l_{j}}A_{i,j}w_{i}+\dee{l_{j}}A_{i}w_{i,j}$.
Substituting $A_{i}w_{i,j}\dee{l_{j}}=\dee{l_{j}}(A_{i}w_{i})_{,j}-\dee{l_{j}}A_{i,j}w_{i}$
in $F_{e}$, renaming indices, and using (\ref{eq:dlxB}), we get
\begin{equation}
\begin{split}F_{e}(\v w) & =I\dee{l_{i}}(A_{i,k}-A_{k,i})w_{k}+I\dee{l_{j}}(A_{i}w_{i})_{,j}\\
 & =I(\dee{\v l}\cross\v B)\cdot\v w+I\dee{\v l}\cdot\nabla(\v A\cdot\v w),
\end{split}
\label{eq:Fe}
\end{equation}
the power of the Lorentz force together with a term that depends on
the gauge. The latter is the derivative of $\v A\cdot\v w$ along
the element, and its sum over a closed loop vanishes. For a single
current element, which is not part of a closed circuit, the energy
is not gauge invariant, and the gauge dependence does not drop out
of the force. This will reappear in Section~\ref{sec:mech} for subbodies
through whose boundary free current flows.

\textbf{Toward distributions of currents and dipoles.} In the rest
of the paper we look for the continuum counterparts of (\ref{eq:Ie}),
(\ref{eq:Im}), (\ref{eq:Fm}) and (\ref{eq:Fe}). We show that, under
weak gauge invariance, the analog of the symmetry of the stress, every
current system localized to subbodies is a superposition of current
elements and magnetic dipoles, that the two parts can be separated
uniquely, and that the purely dipolar systems are exactly those that
are gauge invariant on every subbody.

\section{Stresses, rigid motions and the symmetry of the stress}

\label{sec:stress}

We recall the parts of stress theory that we need. A more detailed
account, with the same notation, is given in {[}GMS26a, Sect.~3{]}.
That the stress exists under Cauchy's hypotheses is Cauchy's classical
theorem (see {[}Nol59{]}, {[}TT60, Sect.~203, p.~543{]}). The distinction
between the traction stress $\v{\gs}$ and the variational stress
$(\v r,\v{\gs})$ was introduced in {[}Seg02{]} for a general manifold
(see also {[}Seg23, Ch.~10{]}).

\textbf{Force systems and the traction stress.} A force system is
a collection $\{F_{\gO}\}$ of linear functionals on virtual velocity
fields $\v w$, one for each regular subbody $\gO$ of a body. Each
$F_{\gO}$ is represented by a body force $\v b$ and a surface force
$\v t_{\gO}$ on $\bdry\gO$, which obviously depends on $\gO$, in
the form 
\begin{equation}
F_{\gO}(\v w)=\int_{\gO}\v b\cdot\v w\dV+\int_{\bdry\gO}\v t_{\gO}\cdot\v w\dA.\label{eq:force}
\end{equation}
By Cauchy's theorem, there is a tensor field $\v{\gs}$, the traction
stress, such that $\v t_{\gO}=\v{\gs}\v{\nor}$, with $\v{\nor}$
the outward unit normal. The traction stress assigns to the velocity
field the vector field $\v{\gs}\trps\v w$, which we call the power
flux vector field. Its flux through $\bdry\gO$ is the power of the
surface forces, since $(\v{\gs}\trps\v w)\cdot\v{\nor}=\v w\cdot\v{\gs}\v{\nor}=\v t_{\gO}\cdot\v w$.
The divergence theorem then gives 
\begin{equation}
F_{\gO}(\v w)=\int_{\gO}\big(\v r\cdot\v w+\gs_{ij}w_{i,j}\big)\dV,\qquad\v r:=\v b+\nabla\cdot\v{\gs},\label{eq:var}
\end{equation}
with $(\nabla\cdot\v{\gs})_{i}=\gs_{ij,j}$. The pair $(\v r,\v{\gs})$
is the variational stress {[}Seg02, Sect.~V{]}, {[}Seg23, Sect.~10.7{]}.
It acts on the first jet $(\v w,\nabla\v w)$, and it represents the
force system on all subbodies by a single field. Note that on a general
manifold, the decompositions $(\v w,\nabla\v w)$ and $(\v r,\v{\gs})$
of the jet and of the variational stress are not invariant. They depend
on the coordinate system used {[}Seg23, Ch.~10{]}. Conversely, the
variational stress determines the representation (\ref{eq:force})
by $\v b=\v r-\nabla\cdot\v{\gs}$ and $\v t_{\gO}=\v{\gs}\v{\nor}$.

\textbf{Rigid motions.} A rigid velocity field is of the form $\v w=\v c+\v{\go}\cross\v x$,
with a translational velocity $\v c$ and an angular velocity $\v{\go}$.
At any given point, its value can be prescribed arbitrarily, by the
choice of $\v c$, and so can its gradient $w_{i,j}=\eps_{ikj}\go_{k}$,
an arbitrary skew-symmetric matrix, by the choice of $\v{\go}$, independently
of the value of $\v w$. By (\ref{eq:var}), 
\begin{equation}
F_{\gO}(\v c+\v{\go}\cross\v x)=\int_{\gO}\v r\cdot(\v c+\v{\go}\cross\v x)\dV+\int_{\gO}\gs_{ij}\eps_{ikj}\go_{k}\dV.\label{eq:rigidpower}
\end{equation}
The first term sees the values of the rigid field, and the second
its gradient, which only the skew-symmetric part of $\v{\gs}$ sees.

In the proofs below we use small balls. We denote by $B_{\gd}$ the
ball of radius $\gd$ about a point $\v x_{0}$, and by $|B_{\gd}|=\frac{4}{3}\pi\gd^{3}$
its volume. If $f$ is continuous, then $\frac{1}{|B_{\gd}|}\int_{B_{\gd}}f\dV\to f(\v x_{0})$
as $\gd\to0$. If, in addition, $\int_{B}f\dV=0$ for every ball $B$,
then $f=0$.
\begin{prop}
\label{prop:rigid} Let $\v b$ be continuous and $\v{\gs}$ of class
$C^{1}$.
\begin{enumerate}
\item[(a)] The power expended by the force for rigid velocity fields depends
only on their values, i.e., there is a continuous vector field $\v g$
such that 
\begin{equation}
F_{\gO}(\v w)=\int_{\gO}\v g\cdot\v w\dV\label{eq:Condition}
\end{equation}
for every $\gO$ and every rigid $\v w$, if and only if $\v{\gs}$
is symmetric. In such a case, $\v g=\v r$.
\item[(b)] The power is invariant under rigid motions, $F_{\gO}(\v w+\v c+\v{\go}\cross\v x)=F_{\gO}(\v w)$
for every $\gO$, $\v w$, $\v c$ and $\v{\go}$, if and only if
$\v{\gs}$ is symmetric and $\v r=0$, i.e., the forces and the moments
on every subbody are balanced.
\end{enumerate}
\end{prop}

\begin{proof}
(a) If $\v{\gs}$ is symmetric, the second term of (\ref{eq:rigidpower})
vanishes, and $\v g=\v r$ satisfies (\ref{eq:Condition}) by (\ref{eq:rigidpower}).
Conversely, assume (\ref{eq:Condition}), let $\v x_{0}$ be a point,
and take $\v c=-\v{\go}\cross\v x_{0}$, so that $\v w=\v{\go}\cross(\v x-\v x_{0})$.
Then, (\ref{eq:Condition}) and (\ref{eq:rigidpower}) on $\gO=B_{\gd}$
imply that 
\[
\int_{B_{\gd}}(\v r-\v g)\cdot\v{\go}\cross(\v x-\v x_{0})\dV+\int_{B_{\gd}}\gs_{ij}\eps_{ikj}\go_{k}\dV=0.
\]
Since $|\v x-\v x_{0}|\leq\gd$ on $B_{\gd}$, the first integral
satisfies 
\[
\Big|\int_{B_{\gd}}(\v r-\v g)\cdot\v{\go}\cross(\v x-\v x_{0})\dV\Big|\leq\gd\,|B_{\gd}|\,|\v{\go}|\max_{B_{\gd}}|\v r-\v g|.
\]
Dividing by $|B_{\gd}|$ and letting $\gd\to0$ gives $\gs_{ij}(\v x_{0})\eps_{ikj}\go_{k}=0$
for every $\v{\go}$, i.e., the skew-symmetric part of $\v{\gs}(\v x_{0})$
vanishes. Then, with $\v{\go}=0$, $\int_{\gO}(\v r-\v g)\cdot\v c\dV=0$
for every $\gO$ and $\v c$, so $\v g=\v r$.

(b) Invariance under rigid motions means, by linearity, that $F_{\gO}(\v w)=0$
for every rigid $\v w$, that is, (\ref{eq:Condition}) with $\v g=0$.
By (a), this holds if and only if $\v{\gs}$ is symmetric and $\v r=\v g=0$.
Finally, $F_{\gO}(\v c)$ is $\v c$ dotted with the total force on
$\gO$, and $F_{\gO}(\v{\go}\cross\v x)$ is $\v{\go}$ dotted with
the total moment about the origin, so the condition says that the
forces and the moments on every subbody are balanced.
\end{proof}

We conclude that the symmetry of the stress is the condition under
which rigid fields contribute to the power only through their values.
In classical continuum mechanics it follows from the balance of moments
in the absence of body couples and couple stresses. Section~\ref{sec:currents}
shows that the corresponding condition in magnetostatics leads to
a skew-symmetric stress.

\section{Current systems}

\label{sec:currents}

We now replace the virtual velocity field $\v w$ by a vector potential
$\v A$ on $\rthree$. As in {[}GMS26a{]}, we may think of $\v A$
either as the potential itself or as a variation of it, assuming that
it can be varied while the current system is held fixed.

\textbf{Current systems.} A current system is a collection $\{I_{\gO}\}$
of linear functionals on vector potentials, one for each regular subbody
$\gO$. Motivated by (\ref{eq:Um}), we interpret $-I_{\gO}(\v A)$
as the potential energy of the currents in $\gO$ under $\v A$, 
\begin{equation}
U_{\gO}(\v A)=-I_{\gO}(\v A).\label{eq:UI}
\end{equation}
In analogy with mechanics, we assume that a functional $I_{\gO}$
is represented by a volume current density $\v j$ and a surface current
density $\v k_{\gO}$ on $\bdry\gO$, in the form 
\begin{equation}
I_{\gO}(\v A)=\int_{\gO}\v j\cdot\v A\dV+\int_{\bdry\gO}\v k_{\gO}\cdot\v A\dA.\label{eq:cursys}
\end{equation}
The surface current $\v k_{\gO}$ depends on the subbody. It is the
current exposed on $\bdry\gO$ when $\gO$ is cut out of the medium.

\textbf{Cauchy's theorem.} Under the usual assumptions of stress theory,
now for a vector-valued surface density acting on $\v A$, Cauchy's
theorem gives a tensor field $\v{\Sigma}$ such that 
\begin{equation}
\v k_{\gO}=\v{\Sigma}\v{\nor},\qquad k_{\gO i}=\Sigma_{ij}\nor_{j}.\label{eq:cauchy}
\end{equation}
This is the magnetostatic analog of the traction stress of mechanics.
It assigns to the potential the vector field $\v{\Sigma}\trps\v A$,
the flux vector field of the current system, which is the analog of
the power flux vector field $\v{\gs}\trps\v w$. Its flux through
$\bdry\gO$ is the contribution of the surface current to $I_{\gO}$,
since $(\v{\Sigma}\trps\v A)\cdot\v{\nor}=\v A\cdot\v{\Sigma}\v{\nor}=\v k_{\gO}\cdot\v A$.
By the divergence theorem, $\int_{\bdry\gO}\v k_{\gO}\cdot\v A\dA=\int_{\gO}(\Sigma_{ij}A_{i})_{,j}\dV=\int_{\gO}(\Sigma_{ij,j}A_{i}+\Sigma_{ij}A_{i,j})\dV$,
and, as in (\ref{eq:var}), 
\begin{equation}
I_{\gO}(\v A)=\int_{\gO}\big(\v r\cdot\v A+\Sigma_{ij}A_{i,j}\big)\dV,\qquad\v r:=\v j+\nabla\cdot\v{\Sigma},\label{eq:repr}
\end{equation}
with $(\nabla\cdot\v{\Sigma})_{i}=\Sigma_{ij,j}$. The pair $(\v r,\v{\Sigma})$
is the magnetostatic analog of the variational stress, acting on the
first jet $(\v A,\nabla\v A)$. It determines the volume current $\v j=\v r-\nabla\cdot\v{\Sigma}$
and the surface currents, and, conversely, it is determined by them.

\textbf{The densities as moments.} The pair $(\v r,\v{\Sigma})$ is
also determined by the values of the current system alone. Let $\v e_{i}$
be the standard basis vectors and let $(\v r,\v{\Sigma})$ be continuous.
For the uniform potential $\v e_{i}$ and the linear potential $(x_{j}-x_{0j})\v e_{i}$,
whose gradient is $\v e_{i}\otimes\v e_{j}$, (\ref{eq:repr}) on
the ball $B_{\gd}$ gives 
\[
I_{B_{\gd}}(\v e_{i})=\int_{B_{\gd}}r_{i}\dV
\]
and 
\[
I_{B_{\gd}}\big((x_{j}-x_{0j})\v e_{i}\big)=\int_{B_{\gd}}r_{i}(x_{j}-x_{0j})\dV+\int_{B_{\gd}}\Sigma_{ij}\dV,
\]
where the first integral on the right is bounded by $\gd\,|B_{\gd}|\max_{B_{\gd}}|\v r|$.
Hence 
\begin{equation}
r_{i}(\v x_{0})=\lim_{\gd\to0}\frac{I_{B_{\gd}}(\v e_{i})}{|B_{\gd}|},\qquad\Sigma_{ij}(\v x_{0})=\lim_{\gd\to0}\frac{I_{B_{\gd}}\big((x_{j}-x_{0j})\v e_{i}\big)}{|B_{\gd}|}.\label{eq:moments}
\end{equation}
Thus, $\v r$ is the density of the net current, and $\v{\Sigma}$
is the density of the first moment of the current, the continuum counterpart
of $\int j_{i}x_{j}\dV$. In particular, $(\v r,\v{\Sigma})$ is unique.

\textbf{Gauge transformations.} A gauge transformation replaces $\v A$
by $\v A+\nabla\gl$ for a twice continuously differentiable function
$\gl$. By (\ref{eq:repr}), it changes the current system by 
\begin{equation}
I_{\gO}(\nabla\gl)=\int_{\gO}\big(\v r\cdot\nabla\gl+\Sigma_{ij}\gl_{,ij}\big)\dV.\label{eq:gaugechange}
\end{equation}
At a point, $\nabla\gl$ can be any vector and the matrix $\gl_{,ij}$
any symmetric matrix, independently. Hence, the gauge fields $\nabla\gl$
play the role of the rigid velocity fields, with the difference that
their gradients are symmetric rather than skew-symmetric. The values
of a gauge field are seen by $\v r$, and its gradient by the symmetric
part of $\v{\Sigma}$.

The change (\ref{eq:gaugechange}) need not vanish. For the current
element (\ref{eq:Ie}) it is $I\dee{\v l}\cdot\nabla\gl(\v y)$. In
general, we can write it in terms of $\v j$ and $\v k_{\gO}$. By
(\ref{eq:cursys}), 
\[
I_{\gO}(\nabla\gl)=\int_{\gO}\v j\cdot\nabla\gl\dV+\int_{\bdry\gO}\v k_{\gO}\cdot\nabla\gl\dA.
\]
Since $\v j\cdot\nabla\gl=\nabla\cdot(\gl\v j)-\gl\,\nabla\cdot\v j$,
the divergence theorem gives 
\begin{equation}
I_{\gO}(\nabla\gl)=-\int_{\gO}\gl\,\nabla\cdot\v j\dV+\int_{\bdry\gO}\gl\,\v j\cdot\v{\nor}\dA+\int_{\bdry\gO}\v k_{\gO}\cdot\nabla\gl\dA.\label{eq:gaugejk}
\end{equation}
The first two terms involve the rate at which the volume current accumulates
charge in $\gO$ and the current that flows out through $\bdry\gO$.
We return to this in Remark~\ref{rem:charge}. We do not require
(\ref{eq:gaugechange}) to vanish, but make the following assumption,
the analog of Proposition~\ref{prop:rigid}(a).
\begin{assumption}[Weak gauge invariance]
\label{ass:weak} The change of a current system under a gauge transformation
depends only on the values of $\nabla\gl$. That is, there is a continuous
vector field $\v g$ such that 
\begin{equation}
I_{\gO}(\nabla\gl)=\int_{\gO}\v g\cdot\nabla\gl\dV\label{eq:weak}
\end{equation}
for every subbody $\gO$ and every function $\gl$. A current system
that satisfies this assumption is called weakly gauge invariant.
\end{assumption}

\begin{prop}
\label{prop:skew} Let $\v j$ be continuous and $\v{\Sigma}$ of
class $C^{1}$. The current system is weakly gauge invariant if and
only if $\v{\Sigma}$ is skew-symmetric, and in such a case, $\v g=\v r$.
\end{prop}

\begin{proof}
If $\v{\Sigma}$ is skew-symmetric, then $\Sigma_{ij}\gl_{,ij}=0$,
since $\gl_{,ij}$ is symmetric, and $\v g=\v r$ satisfies (\ref{eq:weak})
by (\ref{eq:gaugechange}). Conversely, assume (\ref{eq:weak}), let
$\v a$ be a constant symmetric matrix, $\v x_{0}$ a point, and $\gl=\half a_{ij}(x_{i}-x_{0i})(x_{j}-x_{0j})$,
so that $\gl_{,i}=a_{ij}(x_{j}-x_{0j})$ and $\gl_{,ij}=a_{ij}$.
By (\ref{eq:gaugechange}) and (\ref{eq:weak}) on the ball $B_{\gd}$,
\[
\int_{B_{\gd}}(r_{i}-g_{i})\,a_{ij}(x_{j}-x_{0j})\dV+\int_{B_{\gd}}\Sigma_{ij}a_{ij}\dV=0.
\]
The first integral is bounded by $\gd\,|B_{\gd}|\,|\v a|\max_{B_{\gd}}|\v r-\v g|$,
so dividing by $|B_{\gd}|$ and letting $\gd\to0$ gives $\Sigma_{ij}(\v x_{0})a_{ij}=0$
for every symmetric $\v a$, i.e., the symmetric part of $\v{\Sigma}$
vanishes. Then $\int_{\gO}(\v r-\v g)\cdot\nabla\gl\dV=0$ for every
$\gO$ and $\gl$, and $\gl=x_{i}$ gives $\int_{\gO}(r_{i}-g_{i})\dV=0$
for every $\gO$, so $\v g=\v r$ by continuity.
\end{proof}

The reason for the term ``weak gauge invariance'' is the following.
The proof shows that the assumption holds if and only if, at every
point $\v x_{0}$, the density $\v r\cdot\nabla\gl+\Sigma_{ij}\gl_{,ij}$
of (\ref{eq:gaugechange}) vanishes at $\v x_{0}$ for every $\gl$
with $\nabla\gl(\v x_{0})=0$, since $\gl_{,ij}(\v x_{0})$ is then
an arbitrary symmetric matrix. Thus, the energy density is invariant
under every gauge transformation whose gauge field vanishes at the
point, though not, in general, under the others. In the same way,
for a symmetric stress, the power density is invariant under the rigid
velocity fields that vanish at the point, i.e., the rotations about
it.

The physical content of the assumption is seen through (\ref{eq:moments}).
It says that the first moment of the current is skew-symmetric. For
a closed current loop it is, since, by (\ref{eq:loopid}), $I\oint_{C}\xi_{j}\,\dee{\xi_{i}}=\eps_{jik}m_{k}$.
For a pair of antiparallel current elements side by side, which is
not a closed circuit, it is not. Such a pair carries charge to its
four ends, and its energy changes with the second derivatives of $\gl$.
In other words, a steady current circulating in a closed loop has
neither a beginning nor an end, and it does not carry charge from
one place to another. The symmetric part of the first moment measures
such a redistribution of charge. Indeed, for a current density $\v j$
of bounded support and a charge density $\rho$, the conservation
of charge, $\bdry\rho/\bdry t+\nabla\cdot\v j=0$, gives
\[
\int(x_{i}j_{j}+x_{j}j_{i})\dV=\frac{\dee{}}{\dee t}\int x_{i}x_{j}\,\rho\dV.
\]
Hence, for a closed loop, only the skew-symmetric part remains, and
it measures the circulation of the current around an oriented area,
that is, the magnetic dipole moment. The pair of antiparallel current
elements, which carries charge to its ends, has a symmetric first
moment that does not vanish. Thus, the assumption says that the dipoles
of the current system are closed current loops, i.e., magnetic dipoles.
We will see that free current, the zeroth-order part, need not be
closed.

\textbf{The magnetization.} A skew-symmetric $\v{\Sigma}$ has three
independent components, and it is represented by a unique vector field
$\v M$, 
\begin{equation}
\Sigma_{ij}=-\eps_{ijk}M_{k},\qquad M_{k}=-\half\eps_{kij}\Sigma_{ij}.\label{eq:SigmaM}
\end{equation}
The inverse relation follows from $\eps_{kij}\eps_{ijl}=\eps_{ijk}\eps_{ijl}=2\gd_{kl}$,
which is (\ref{eq:epsdelta}) contracted once more. For example, $\Sigma_{12}=-M_{3}$,
$\Sigma_{23}=-M_{1}$ and $\Sigma_{31}=-M_{2}$. We can now express
the terms of (\ref{eq:repr}) and (\ref{eq:cauchy}) by $\v M$ as
\begin{equation}
\begin{split}\Sigma_{ij}A_{i,j} & =\eps_{kji}M_{k}A_{i,j}=\v M\cdot\v B,\\
k_{\gO i} & =\Sigma_{ij}\nor_{j}=-\eps_{ijk}\nor_{j}M_{k}=-(\v{\nor}\cross\v M)_{i}=(\v M\cross\v{\nor})_{i},\\
(\nabla\cdot\v{\Sigma})_{i} & =\Sigma_{ij,j}=-\eps_{ijk}M_{k,j}=-(\nabla\cross\v M)_{i},
\end{split}
\label{eq:SigmaA}
\end{equation}
where, in the first line, we used $B_{k}=\eps_{kij}A_{j,i}=\eps_{kji}A_{i,j}$
and $-\eps_{ijk}=\eps_{kji}$. In the same way, the flux vector field
is $(\v{\Sigma}\trps\v A)_{j}=\Sigma_{ij}A_{i}=-\eps_{jki}A_{i}M_{k}=(\v A\cross\v M)_{j}$,
i.e., 
\begin{equation}
\v{\Sigma}\trps\v A=\v A\cross\v M,\qquad(\v A\cross\v M)\cdot\v{\nor}=\v A\cdot(\v M\cross\v{\nor})=\v k_{\gO}\cdot\v A,\label{eq:fluxAM}
\end{equation}
as it should be. Hence, by Proposition~\ref{prop:skew}, a current
system is weakly gauge invariant if and only if its flux vector field
has the form $\v A\cross\v M$ for some vector field $\v M$. This
is the magnetostatic counterpart of the requirement that the power
flux vector field $\v{\gs}\trps\v w$ be given by a symmetric stress.

\textbf{Free and bound currents.} For a weakly gauge-invariant current
system, we write $\v j_{f}:=\v r$. By (\ref{eq:repr}) and (\ref{eq:SigmaA}),
every such current system takes the form 
\begin{equation}
\begin{split}I_{\gO}(\v A) & =\int_{\gO}\big(\v j_{f}\cdot\v A+\v M\cdot\v B\big)\dV,\\
U_{\gO}(\v A) & =-\int_{\gO}\v j_{f}\cdot\v A\dV-\int_{\gO}\v M\cdot\v B\dV,
\end{split}
\label{eq:repM}
\end{equation}
and, since $\v j=\v r-\nabla\cdot\v{\Sigma}$, 
\begin{equation}
\v j=\v j_{f}+\v j_{b},\qquad\v j_{b}:=\nabla\cross\v M,\qquad\v k_{\gO}=\v M\cross\v{\nor}.\label{eq:bound}
\end{equation}
The current of every subbody splits uniquely into two parts. The first
is the free current, with density $\v j_{f}$. Its energy is the superposition
of current elements (\ref{eq:Ie}) with $I\dee{\v l}=\v j_{f}\dV$.
The second is the current bound in a distribution of magnetic dipoles
with density $\v M$, consisting of the bound volume current $\v j_{b}$
and the bound surface current $\v k_{\gO}$. Its energy is the superposition
of dipoles (\ref{eq:Im}) with $\v m=\v M\dV$. In the terminology
of {[}Jac99, Zan13{]}, $\v j_{f}$ is the free current density and
$\v M$ is the magnetization. The division of the current into free
and bound parts has been regarded as partly arbitrary. For instance,
it is stated in {[}PP62, p.~28{]} that the division of the charge
into free and bound charge ``is to a certain extent arbitrary'',
and a similar remark on currents is made in {[}PP62, p.~139{]}. This
is indeed the case if one considers only the total current $\v j=\v j_{f}+\nabla\cross\v M$,
or the energy of the body as a whole. The total current does not change
if $\v j_{f}$ is replaced by $\v j_{f}+\nabla\cross\v h$ and $\v M$
by $\v M-\v h$, for any vector field $\v h$, and, if $\v h$ vanishes
on the boundary of the body, neither does the energy of the body.
Here, the split is unique, because a current system assigns an energy
to every subbody, and the replacement changes the surface current
$\v M\cross\v{\nor}$ exposed on the boundaries of the subbodies,
and with it their energies (see also Remark~\ref{rem:Mf}).

\textbf{The densities as moments, again.} The densities have a direct
interpretation. By (\ref{eq:moments}), $\v j_{f}$ is the density
of the net current. For $\v M$, we use the potential $\half\v e_{k}\cross(\v x-\v x_{0})$.
For a constant vector $\v c$, $(\v c\cross\v x)_{i}=\eps_{ipq}c_{p}x_{q}$,
and, by (\ref{eq:epsdelta}), 
\begin{equation}
\begin{split}\big(\nabla\cross(\v c\cross\v x)\big)_{l} & =\eps_{lmi}\eps_{ipq}c_{p}\gd_{qm}=\eps_{ilm}\eps_{ipm}c_{p}\\
 & =(\gd_{lp}\gd_{mm}-\gd_{lm}\gd_{mp})c_{p}=3c_{l}-c_{l}=2c_{l},
\end{split}
\label{eq:curlcx}
\end{equation}
so that $\nabla\cross\big(\half\v c\cross(\v x-\v x_{0})\big)=\v c$,
since a constant $\v x_{0}$ does not affect the curl. With $\v c=\v e_{k}$
and (\ref{eq:repM}) on $B_{\gd}$, 
\[
I_{B_{\gd}}\big(\half\v e_{k}\cross(\v x-\v x_{0})\big)=\half\int_{B_{\gd}}\v j_{f}\cdot\big(\v e_{k}\cross(\v x-\v x_{0})\big)\dV+\int_{B_{\gd}}M_{k}\dV.
\]
The first integral is bounded by $\half\gd\,|B_{\gd}|\max_{B_{\gd}}|\v j_{f}|$,
so dividing by $|B_{\gd}|$ and letting $\gd\to0$, 
\begin{equation}
M_{k}(\v x_{0})=\lim_{\gd\to0}\frac{I_{B_{\gd}}\big(\half\v e_{k}\cross(\v x-\v x_{0})\big)}{|B_{\gd}|}.\label{eq:Mmoment}
\end{equation}
Why does this potential measure the magnetic moment? By (\ref{eq:curlcx}),
$\half\v B_{0}\cross(\v x-\v x_{0})$ is a potential of the uniform
field $\v B_{0}$. For a loop, $(\v B_{0}\cross\v{\xi})\cdot\dee{\v{\xi}}=\eps_{ijk}B_{0j}\xi_{k}\,\dee{\xi_{i}}$,
so that, by (\ref{eq:loopid}) and $\eps_{ijk}\eps_{kil}=\eps_{kij}\eps_{kil}=2\gd_{jl}$,
\[
I\oint_{C}\half(\v B_{0}\cross\v{\xi})\cdot\dee{\v{\xi}}=\half I\eps_{ijk}B_{0j}\eps_{kil}a_{l}=I\,\v B_{0}\cdot\v a=\v m\cdot\v B_{0}.
\]
This is exactly $-U_{m}$ of (\ref{eq:Um}), now without approximation,
since the field is uniform. For the subbody $\gO$, (\ref{eq:repM}),
the identity $\v j_{f}\cdot(\v B_{0}\cross\v y)=\v B_{0}\cdot(\v y\cross\v j_{f})$,
and $\nabla\cross\big(\half\v B_{0}\cross(\v x-\v x_{0})\big)=\v B_{0}$
give 
\begin{equation}
\begin{split}I_{\gO}\big(\half\v B_{0}\cross(\v x-\v x_{0})\big) & =\half\int_{\gO}\v B_{0}\cdot\big((\v x-\v x_{0})\cross\v j_{f}\big)\dV+\int_{\gO}\v M\cdot\v B_{0}\dV\\
 & =\v B_{0}\cdot\v m_{\gO},\\
\v m_{\gO} & :=\half\int_{\gO}(\v x-\v x_{0})\cross\v j_{f}\dV+\int_{\gO}\v M\dV.
\end{split}
\label{eq:mOmega}
\end{equation}
Thus, $\v M$ is the density of magnetic moment, and $\v m_{\gO}$
is the familiar magnetic moment of a current distribution {[}Jac99{]}
plus that of the magnetization. In general, $\v m_{\gO}$ depends
on $\v x_{0}$, through $\half\v x_{0}\cross\int_{\gO}\v j_{f}\dV$.
It does not when the net free current of $\gO$ vanishes.

\textbf{The analogy.} Table~\ref{tab:analogy} summarizes the correspondence
with Section~\ref{sec:stress}. The vector bundle of generalized
velocities is the same as in continuum mechanics, but the role of
the rigid fields is taken by the gauge fields. Their gradients are
symmetric, so the condition of Proposition~\ref{prop:rigid}(a) turns
into skew-symmetry, and the stress is represented by a vector field.

\begin{table}[ht]
\begin{tabular}{ll}
\toprule 
Continuum mechanics & Magnetostatics\tabularnewline
\midrule 
virtual velocity $\v w$ & vector potential $\v A$\tabularnewline
rigid fields $\v c+\v{\go}\cross\v x$, $\nabla\v w$ skew & gauge fields $\nabla\gl$, $\nabla\nabla\gl$ symmetric\tabularnewline
body force $\v b$ & volume current $\v j$\tabularnewline
unbalanced force $\v r=\v b+\nabla\cdot\v{\gs}$ & free current $\v j_{f}=\v j+\nabla\cdot\v{\Sigma}$\tabularnewline
surface traction $\v t_{\gO}=\v{\gs}\v{\nor}$ & surface current $\v k_{\gO}=\v{\Sigma}\v{\nor}$\tabularnewline
traction stress $\v{\gs}$ & $\v{\Sigma}$\tabularnewline
power flux vector field $\v{\gs}\trps\v w$ & flux vector field $\v{\Sigma}\trps\v A=\v A\cross\v M$\tabularnewline
variational stress $(\v r,\v{\gs})$ & $(\v j_{f},\v{\Sigma})$\tabularnewline
virtual power $F_{\gO}(\v w)$ & minus the energy $I_{\gO}(\v A)=-U_{\gO}(\v A)$\tabularnewline
symmetry $\v{\gs}=\v{\gs}\trps$ & skew-symmetry $\v{\Sigma}=-\v{\Sigma}\trps$, i.e., $\v M$\tabularnewline
stress power $\int\gs_{ij}w_{i,j}\dV$ & minus the energy of dipoles $\int\v M\cdot\v B\dV$\tabularnewline
sum of forces on $\gO$ & net current of $\gO$, $I_{\gO}(\v e_{i})$\tabularnewline
balance $\v b+\nabla\cdot\v{\gs}=0$ & $\v j=\nabla\cross\v M$\tabularnewline
every subbody in equilibrium & no net current in any subbody\tabularnewline
\bottomrule
\end{tabular}\vspace{3mm}
 \caption{The analogy between continuum mechanics and magnetostatics.}
\label{tab:analogy}
\end{table}

\section{Gauge invariance and closed currents}

\label{sec:gauge}

\subsection{The main result}

For a weakly gauge-invariant current system, (\ref{eq:repM}) gives
\begin{equation}
I_{\gO}(\v A+\nabla\gl)-I_{\gO}(\v A)=\int_{\gO}\v j_{f}\cdot\nabla\gl\dV,\label{eq:gaugeU}
\end{equation}
since $\nabla\cross\nabla\gl=0$. If $\v A$ is a variation of the
potential, this change is physical, as for the current element (\ref{eq:Ie}).
Thus, gauge invariance is a property of a special class of current
systems, just as equilibrium is a property of a special class of force
systems.
\begin{defn}
\label{def:gauge} A current system $\{I_{\gO}\}$ is gauge invariant
if $I_{\gO}(\v A+\nabla\gl)=I_{\gO}(\v A)$ for every subbody $\gO$,
every vector potential $\v A$, and every function $\gl$.
\end{defn}

\begin{prop}
\label{prop:main} For a weakly gauge-invariant current system with
continuous $\v j$ and $C^{1}$ field $\v{\Sigma}$, the following
are equivalent.
\begin{enumerate}
\item[(i)] No subbody carries a net current: $\int_{\gO}\v j\dV+\int_{\bdry\gO}\v k_{\gO}\dA=0$
for every $\gO$.
\item[(ii)] The current system is gauge invariant.
\item[(iii)] $\v j_{f}=0$, i.e., $\v j=\nabla\cross\v M$.
\item[(iv)] $I_{\gO}(\v A)=\int_{\gO}\v M\cdot\v B\dV$, i.e., $U_{\gO}(\v A)=-\int_{\gO}\v M\cdot\v B\dV$,
for every $\gO$ and $\v A$.
\end{enumerate}
\end{prop}

\begin{proof}
The proof follows that of {[}GMS26a, Prop.~5.2{]}. By (\ref{eq:cursys}),
the $i$-th component of the left side of (i) is $I_{\gO}(\v e_{i})$,
and $\v e_{i}=\nabla x_{i}$ is a gauge field. So, if the current
system is gauge invariant, $I_{\gO}(\v e_{i})=I_{\gO}(0+\nabla x_{i})=I_{\gO}(0)=0$,
and (ii)$\Rightarrow$(i). By (\ref{eq:repM}), $I_{\gO}(\v e_{i})=\int_{\gO}j_{f\,i}\dV$,
which vanishes for every $\gO$ if and only if $\v j_{f}=0$, by continuity;
so (i)$\Leftrightarrow$(iii). (iii)$\Rightarrow$(iv) by (\ref{eq:repM}),
and (iv)$\Rightarrow$(ii) since $\nabla\cross\nabla\gl=0$.
\end{proof}

We call the property of Definition~\ref{def:gauge} simply gauge
invariance, as in {[}GMS26a{]}, and not strong gauge invariance. In
electrostatics, every charge system is weakly gauge invariant, since
a scalar gauge transformation has no derivatives to act on, and gauge
invariance there means exactly what it means here. Without weak gauge
invariance, the same argument, with (\ref{eq:gaugechange}) and Proposition~\ref{prop:skew},
shows that a current system is gauge invariant if and only if $\v{\Sigma}$
is skew-symmetric and $\v r=0$. This is the analog of Proposition~\ref{prop:rigid}(b):
invariance under all rigid motions requires both the symmetry of the
stress and the balance of forces. Hence, gauge invariance implies
weak gauge invariance, just as invariance under rigid motions implies
the symmetry of the stress.

Thus, gauge invariance of the energy is to magnetostatics what the
balance of forces on every subbody is to continuum mechanics. For
weakly gauge-invariant systems, it singles out the current systems
without free current, and the representation $-\int\v M\cdot\v B\dV$
of their energy is the analog of the representation of the power of
an equilibrated force system by $\int\gs_{ij}w_{i,j}\dV$. A body
carrying free current is the analog of a body acted upon by unbalanced
forces.

\subsection{Consequences}

\textbf{The energy depends only on the magnetic field.} For a system
without free current, the energy depends on $\v A$ only through $\v B$,
as does the energy of a single dipole (property (b) of Section~\ref{sec:dipole-functional}).
In general, it also depends on $\v A$ itself, through the free current.

\textbf{Gauge invariance on a single body.} The energy of a single
body $\gO$ may be gauge invariant even if $\v j_{f}\neq0$. By (\ref{eq:gaugeU}),
$\v j_{f}\cdot\nabla\gl=\nabla\cdot(\gl\v j_{f})-\gl\,\nabla\cdot\v j_{f}$,
and the divergence theorem, 
\begin{equation}
\int_{\gO}\v j_{f}\cdot\nabla\gl\dV=\int_{\bdry\gO}\gl\,\v j_{f}\cdot\v{\nor}\dA-\int_{\gO}\gl\,\nabla\cdot\v j_{f}\dV.\label{eq:gaugebody}
\end{equation}
This vanishes for every $\gl$ if and only if $\nabla\cdot\v j_{f}=0$
in $\gO$ and $\v j_{f}\cdot\v{\nor}=0$ on $\bdry\gO$, that is,
if the free current closes within $\gO$. A subbody $\gO'$ cut out
of such a body, such that $\v j_{f}$ flows through $\bdry\gO'$,
does not have a gauge-invariant energy. This is the analog of a body
in equilibrium whose parts are not, and of a neutral body containing
free charges of both signs.

\textbf{Bound currents.} For a magnetic dipole distribution, $\v M$
determines the current system completely, namely, the bound volume
current $\nabla\cross\v M$ and the bound surface current $\v M\cross\v{\nor}$,
as in {[}Jac99, Sect.~5.8{]}, {[}Zan13{]}, with the correct signs.
They can be read off from a model of small loops. For a uniform $\v M$,
the currents of neighboring loops cancel in the interior, and only
the surface current $\v M\cross\v{\nor}$ remains. If $\v M=M_{3}(x_{1})\v e_{3}$
increases with $x_{1}$, the loops on the right are stronger, and
the net current between them flows in the direction of $-\v e_{2}$,
which is $\nabla\cross\v M=-M_{3}'\v e_{2}$. The bound current is
divergence free, and, as a result, it carries no charge out of any
subbody. Indeed, the total bound current through the boundary of every
subbody, $\int_{\bdry\gO}(\nabla\cross\v M)\cdot\v{\nor}\dA$, vanishes.

\textbf{The meaning of the surface current.} The surface current $\v M\cross\v{\nor}$
is defined for every subbody, not only for the boundary of the medium.
On an internal surface it is the current exposed when the subbody
is cut out, namely, the halves of the current loops severed by the
cut. This is the exact analog of the traction on an internal surface
in continuum mechanics.
\begin{rem}[Free current as a curl]
\label{rem:Mf} Let $\v j_{f}$ be defined on all of $\rthree$,
divergence free, in the sense of distributions, and of bounded support.
Then there is a vector field $\v M_{f}$ with 
\begin{equation}
\nabla\cross\v M_{f}=\v j_{f},\label{eq:Mf}
\end{equation}
for example the field given by the Biot\textendash Savart law, 
\begin{equation}
\v M_{f}(\v x)=\frac{1}{4\pi}\int\v j_{f}(\v y)\cross\frac{\v x-\v y}{|\v x-\v y|^{3}}\,\dee{V_{y}}.\label{eq:BS}
\end{equation}
In such a case, the vector field $\v M_{f}$ is unique up to a gradient.
The Biot\textendash Savart field above is the unique divergence-free
choice that vanishes at infinity. No condition on the topology of
$\gO$ is needed here, since $\v j_{f}$ is defined on all of $\rthree$,
where every divergence-free field of bounded support is a curl. By
the identity $\nabla\cdot(\v M_{f}\cross\v A)=\v A\cdot\nabla\cross\v M_{f}-\v M_{f}\cdot\nabla\cross\v A$
and the divergence theorem, with $(\v M_{f}\cross\v A)\cdot\v{\nor}=-(\v M_{f}\cross\v{\nor})\cdot\v A$,
\begin{equation}
\int_{\gO}\v j_{f}\cdot\v A\dV=\int_{\gO}\v M_{f}\cdot\v B\dV-\int_{\bdry\gO}(\v M_{f}\cross\v{\nor})\cdot\v A\dA.\label{eq:freeasM}
\end{equation}
Thus, the free current has the energy of the dipole distribution $\v M_{f}$,
whose bound volume current is $\nabla\cross\v M_{f}=\v j_{f}$, stripped
of its bound surface current $\v M_{f}\cross\v{\nor}$. Indeed, by
(\ref{eq:bound}), the dipole distribution $\v M_{f}$, viewed as
a current system, has the bound volume current $\nabla\cross\v M_{f}=\v j_{f}$
and the bound surface current $\v M_{f}\cross\v{\nor}$ on $\bdry\gO$.
By (\ref{eq:cursys}) and (\ref{eq:repM}), its energy is therefore
\[
-\int_{\gO}\v M_{f}\cdot\v B\dV=-\int_{\gO}\v j_{f}\cdot\v A\dV-\int_{\bdry\gO}(\v M_{f}\cross\v{\nor})\cdot\v A\dA,
\]
which is (\ref{eq:freeasM}). This is the continuum form of Ampère's
equivalence between a current loop and a magnetic shell, and it is
the analog of {[}GMS26a, Remark~5.6{]}. Writing $\v j_{f}=\nabla\cross\v M_{f}$
does not turn free current into dipoles. The field $\v M_{f}$ is
not determined by the current system, and the free current is the
part of the current that exposes no current on cuts. The first term
on the right of (\ref{eq:freeasM}) is gauge invariant, so the whole
gauge dependence of the energy of the free current sits in the boundary
term, the energy of the missing surface current. It vanishes if, in
addition, $\v j_{f}\cdot\v{\nor}=0$ on $\bdry\gO$, in agreement
with (\ref{eq:gaugebody}).
\end{rem}

\begin{rem}[Conservation of charge]
\label{rem:charge} For every subbody $\gO$, the balance of charge
reads 
\begin{equation}
\frac{\dee{}}{\dee t}\int_{\gO}\rho\dV=-\int_{\bdry\gO}\v j\cdot\v{\nor}\dA,\label{eq:chargebalance}
\end{equation}
where $\rho$ is the charge density and $\v j$ is the total current
density, free and bound. That is, the charge in $\gO$ changes only
through the current that flows through $\bdry\gO$. Since this holds
for every subbody, the divergence theorem gives the local form 
\begin{equation}
\bdry_{t}\rho+\nabla\cdot\v j=0.\label{eq:chargelocal}
\end{equation}
In magnetostatics, the charge density does not change in time. Hence,
conservation of charge is the condition $\nabla\cdot\v j=0$ on the
total current. For a weakly gauge-invariant current system, $\nabla\cdot\v j_{b}=\nabla\cdot(\nabla\cross\v M)=0$.
Thus, the bound current always conserves charge, and the condition
reduces to $\nabla\cdot\v j_{f}=0$, a condition on the free current
alone.

A current system need not satisfy this condition. Consider a current
element $I\dee{\v l}$ at a point $\v y$ in the interior of $\gO$,
i.e., a short piece of wire both ends of which lie in $\gO$. Its
current density is $\v j=I\dee{\v l}\,\gd_{\v y}$, and, in the sense
of distributions, 
\[
\int\gl\,\nabla\cdot\v j\dV=-\int\v j\cdot\nabla\gl\dV=-I\dee{\v l}\cdot\nabla\gl(\v y)
\]
for every smooth function $\gl$ with compact support. It follows
that $\nabla\cdot\v j=I\dee{\v l}\cdot\nabla\gd_{\v y}\neq0$. Physically,
current flows into the element at one end and out of it at the other,
and there is nothing to supply or remove the charge. Hence, by (\ref{eq:chargelocal}),
charge accumulates at the head of the element and is depleted at its
tail. On the other hand, a wire segment that runs from one point of
$\bdry\gO$ to another, carrying a constant current, conserves charge
in $\gO$. Its current, however, crosses $\bdry\gO$, as in Example~\ref{ex:wire}.

Gauge invariance is closely related to conservation of charge, and
the relation becomes precise if we restrict the gauge functions. Let
$\gl$ vanish in a neighborhood of $\bdry\gO$. Then the boundary
terms of (\ref{eq:gaugejk}) drop out, and 
\begin{equation}
I_{\gO}(\nabla\gl)=-\int_{\gO}\gl\,\nabla\cdot\v j\dV,\label{eq:gaugecharge}
\end{equation}
for every current system, weakly gauge invariant or not. It follows
that the energy of $\gO$ is invariant under all gauge transformations
that vanish in a neighborhood of $\bdry\gO$ if and only if $\nabla\cdot\v j=0$
in $\gO$, that is, if and only if the currents conserve charge in
$\gO$. This is the static form of the familiar correspondence between
gauge invariance and conservation of charge. For the current element
above, $I_{\gO}(\nabla\gl)=I\dee{\v l}\cdot\nabla\gl(\v y)$, which
does not vanish in general, even if $\gl$ vanishes near $\bdry\gO$.
For a wire segment that runs from $\v y_{1}\in\bdry\gO$ to $\v y_{2}\in\bdry\gO$,
(\ref{eq:gaugejk}) gives $I_{\gO}(\nabla\gl)=I\big(\gl(\v y_{2})-\gl(\v y_{1})\big)$,
which vanishes when $\gl$ vanishes near $\bdry\gO$, but not in general.

Conservation of charge is also exactly the hypothesis of Remark~\ref{rem:Mf}.
If $\v j_{f}$ is defined on all of $\rthree$, $\nabla\cdot\v j_{f}=0$,
and $\v j_{f}$ has bounded support, then $\v j_{f}=\nabla\cross\v M_{f}$.
Conversely, $\v j_{f}=\nabla\cross\v M_{f}$ implies $\nabla\cdot\v j_{f}=0$.
By (\ref{eq:freeasM}), the energy of a charge-conserving free current
in $\gO$ is the energy of the dipole distribution $\v M_{f}$, which
is gauge invariant, minus the energy of the surface current $\v M_{f}\cross\v{\nor}$
on $\bdry\gO$. A gauge transformation that vanishes near $\bdry\gO$
does not change the latter. This is another way of seeing why such
gauge transformations leave the energy invariant. Gauge invariance
on every subbody, Definition~\ref{def:gauge}, is much stronger.
By (\ref{eq:gaugebody}), it also rules out free current that crosses
the boundary of any subbody, and, by Proposition~\ref{prop:main},
it rules out free current altogether.
\end{rem}

\subsection{Three levels of gauge invariance}

\label{sec:levels}

We now have three conditions on the energy of a current system, each
stronger than the previous one. For a current system with continuous
$\v j$ and $C^{1}$ field $\v{\Sigma}$:
\begin{enumerate}
\item[(1)] Weak gauge invariance (Assumption~\ref{ass:weak}) holds if and
only if $\v{\Sigma}$ is skew-symmetric, i.e., if and only if $\v{\Sigma}$
is represented by a magnetization $\v M$ (Proposition~\ref{prop:skew}).
Physically, the dipoles of the current system are closed current loops.
This is the analog of the symmetry of the stress.
\item[(2)] For a weakly gauge-invariant current system, invariance under all
gauge transformations that vanish near the boundary of the subbody
holds for every subbody if and only if, in addition, $\nabla\cdot\v j_{f}=0$
(Remark~\ref{rem:charge}). Physically, the currents conserve charge.
This is the natural setting of magnetostatics.
\item[(3)] For a weakly gauge-invariant current system, gauge invariance (Definition~\ref{def:gauge})
holds if and only if, in addition, $\v j_{f}=0$ (Proposition~\ref{prop:main}).
Physically, there is no free current. This is the analog of the balance
of forces.
\end{enumerate}
In the rest of the paper, we work at the second level.

\section{Mechanical forces on embedded current distributions}

\label{sec:mech}

\subsection{Embedded current distributions}

Let $\gO$ be a material body and $\dis\colon\reals\times\gO\to\rthree$
a virtual motion such that $\dis_{0}=\mathrm{Id}$ and the configurations
$\dis_{t}:=\dis(t,\cdot)$ are orientation-preserving embeddings,
with deformation gradient $\v F_{t}=\nabla\dis_{t}$ and Jacobian
$J_{t}=\det\v F_{t}>0$. As in Section~\ref{sec:dipole-functional},
let the virtual velocity field be given by $\v w=\bdry_{t}\dis_{t}|_{t=0}$.
The potential $\v A$ is held fixed in space.

We say that the current distribution $(\v j_{f},\v M)$ is embedded
in the body if, as in Section~\ref{sec:dipole-functional}, each
free current element is carried as a material line element with its
current, and each magnetic dipole is carried as a material area element
with its current. For the free current element $\v j_{f}\dV$ this
means $\v j_{f\,t}\dee{V_{t}}=\v F_{t}\v j_{f}\dV$, and for the dipole
moment $\v M\dV$ it means $\v M_{t}\dee{V_{t}}=J_{t}\v F^{-\mathsf{T}}_{t}\v M\dV$.
Since $\dee{V_{t}}=J_{t}\dV$, 
\begin{equation}
\v j_{f\,t}=\frac{1}{J_{t}}\v F_{t}\v j_{f},\qquad\v M_{t}=\v F^{-\mathsf{T}}_{t}\v M.\label{eq:convect}
\end{equation}
The free current is carried by the Piola transformation, like the
polarization in {[}GMS26a{]}. By Nanson's formula, $\v{\nor}_{t}\dee{A_{t}}=J_{t}\v F^{-\mathsf{T}}_{t}\v{\nor}\dA$,
so that 
\[
\v j_{f\,t}\cdot\v{\nor}_{t}\dee{A_{t}}=\frac{1}{J_{t}}\v F_{t}\v j_{f}\cdot J_{t}\v F^{-\mathsf{T}}_{t}\v{\nor}\dA=\v j_{f}\cdot\v{\nor}\dA,
\]
that is, the free current through every material surface is conserved.
The magnetization is carried as a covector, so $\v M_{t}\cdot\v F_{t}\dee{\v x}=\v M\cdot\v F^{-1}_{t}\v F_{t}\dee{\v x}=\v M\cdot\dee{\v x}$.
This implies that the line integral of $\v M$ along every material
curve is conserved. For a closed material curve, it is the circulation
of $\v M$, and, by Stokes' theorem, the bound current $\int(\nabla\cross\v M)\cdot\v{\nor}\dA$
through every material surface is conserved as well.

What is embedded is the current flow, not the charge carriers. In
a wire attached to the body, the charge carriers move along the wire,
relative to the material. What the embedding assumption says is that
the streamlines of the free current are material curves, which move
and deform with the body, and that the current through every material
cross-section is not affected by the motion. This is what we assumed
for the single loop and the single current element in Section~\ref{sec:dipole}.
The assumption rules out conduction in a deforming conductor, where
the current redistributes itself relative to the material as the body
deforms. Nor do we consider the electromotive forces induced by the
motion. As in Section~\ref{sec:dipole}, the currents are held fixed
by their sources.

\textbf{Conservation of charge.} From now on, we postulate that the
free current conserves charge, $\nabla\cdot\v j_{f}=0$, the second
level of Section~\ref{sec:levels}. In magnetostatics, this is not
an additional restriction. By Remark~\ref{rem:charge}, it follows
from the balance of charge when the charge density does not change
in time, and Maxwell's equations require it as well (Section~\ref{sec:concl}).
It excludes isolated current elements, which are not static anyway.
We will point out where the assumption is used.

\subsection{The force functional}

The energy (\ref{eq:repM}) of the convected distribution at time
$t$ is 
\begin{equation}
\begin{split}U_{t} & =-\int_{\dis_{t}(\gO)}\big(\v j_{f\,t}\cdot\v A+\v M_{t}\cdot\v B\big)\dee{V_{t}}\\
 & =-\int_{\gO}\dis_{t\,i,j}(\v x)\,j_{f\,j}(\v x)\,A_{i}(\dis_{t}(\v x))\dV\\
 & \qquad-\int_{\gO}J_{t}(\v x)\,(\v F^{-\mathsf{T}}_{t}(\v x))_{ij}M_{j}(\v x)\,B_{i}(\dis_{t}(\v x))\dV.
\end{split}
\label{eq:Ut}
\end{equation}
To get the second line, we changed variables to $\v x\in\gO$, with
$\dee{V_{t}}=J_{t}\dV$, and used (\ref{eq:convect}): $\v j_{f\,t}\dee{V_{t}}=\v F_{t}\v j_{f}\dV$
and $\v M_{t}\dee{V_{t}}=J_{t}\v F^{-\mathsf{T}}_{t}\v M\dV$. Now
we differentiate at $t=0$ under the integral sign, exactly as in
Section~\ref{sec:dipole-functional}. We use three identities: (\ref{eq:Fdot})
for the derivative of $\dis_{t\,i,j}$, which is $w_{i,j}$ at $t=0$;
(\ref{eq:JFTdot}) for the derivative of $J_{t}\v F^{-\mathsf{T}}_{t}$;
and the chain rule, $\bdry_{t}A_{i}(\dis_{t}(\v x))|_{t=0}=A_{i,k}w_{k}$,
and similarly for $\v B$. At $t=0$, $\dis_{0\,i,j}=\gd_{ij}$ and
$J_{0}\v F^{-\mathsf{T}}_{0}=\v I$, so 
\[
\begin{split}\frac{\dee{}}{\dee t}U_{t}\Big|_{t=0}= & -\int_{\gO}\big(w_{i,j}j_{f\,j}A_{i}+j_{f\,i}A_{i,k}w_{k}\big)\dV\\
 & -\int_{\gO}\big((w_{k,k}\gd_{ij}-w_{j,i})M_{j}B_{i}+M_{i}B_{i,k}w_{k}\big)\dV.
\end{split}
\]
As in Section~\ref{sec:dipole}, we define the force functional as
$F_{\gO}(\v w):=-\dee{U_{t}}/\dee t|_{t=0}$. Renaming $i\leftrightarrow j$
in the term $w_{j,i}M_{j}B_{i}$, 
\begin{equation}
\begin{split}F_{\gO}(\v w)= & \int_{\gO}\big(j_{f\,j}A_{i}w_{i,j}+j_{f\,i}A_{i,k}w_{k}\big)\dV\\
 & +\int_{\gO}\big(M_{i}B_{i,k}w_{k}+(\v M\cdot\v B)w_{k,k}-M_{i}B_{j}w_{i,j}\big)\dV.
\end{split}
\label{eq:F1}
\end{equation}
The second integral is the continuum form of (\ref{eq:Fm}). The first
is the continuum form of the expression for $F_{e}$ in Section~\ref{sec:dipole-functional},
and we rearrange it exactly as we did there. Since $(\v j_{f}\cross\v B)_{k}=j_{f\,i}(A_{i,k}-A_{k,i})$,
by (\ref{eq:dlxB}) with $\dee{\v l}$ replaced by $\v j_{f}$, and
$j_{f\,j}(A_{i}w_{i})_{,j}=j_{f\,j}A_{i,j}w_{i}+j_{f\,j}A_{i}w_{i,j}$,
we have 
\[
j_{f\,j}A_{i}w_{i,j}+j_{f\,i}A_{i,k}w_{k}=(\v j_{f}\cross\v B)\cdot\v w+\v j_{f}\cdot\nabla(\v A\cdot\v w).
\]
Thus, 
\begin{equation}
\begin{split}F_{\gO}(\v w)= & \int_{\gO}\big(\v j_{f}\cross\v B+(\nabla\v B)\trps\v M\big)\cdot\v w\dV\\
 & +\int_{\gO}\big((\v M\cdot\v B)\v I-\v M\otimes\v B\big):\nabla\v w\dV\\
 & +\int_{\gO}\v j_{f}\cdot\nabla(\v A\cdot\v w)\dV.
\end{split}
\label{eq:F2}
\end{equation}
Comparing with (\ref{eq:Fm}) and (\ref{eq:Fe}), the continuum theory
may be viewed as the analog of a continuous superposition of embedded
current elements and magnetic dipoles, both for the energy and for
the force.
\begin{prop}
\label{prop:forcegauge} Let $\gO$ be a particular subbody. Under
a gauge transformation $\v A\mapsto\v A+\nabla\gl$, the force functional
(\ref{eq:F2}) changes by 
\begin{equation}
\gd F_{\gO}(\v w)=\int_{\gO}\v j_{f}\cdot\nabla(\v w\cdot\nabla\gl)\dV,\label{eq:dF}
\end{equation}
which, by (\ref{eq:gaugebody}) and the conservation of charge, may
be written as 
\[
\gd F_{\gO}(\v w)=\int_{\bdry\gO}(\v w\cdot\nabla\gl)\,\v j_{f}\cdot\v{\nor}\dA.
\]
Consequently:
\begin{enumerate}
\item[(a)] The force on the magnetization, the part of $F_{\gO}$ involving
$\v M$, is gauge invariant.
\item[(b)] $F_{\gO}(\v w)$ is gauge invariant for every $\v w$ if and only
if $\v j_{f}\cdot\v{\nor}=0$ on $\bdry\gO$, and it is gauge invariant
for every $\v w$ that vanishes on $\bdry\gO$.
\end{enumerate}
\end{prop}

\begin{proof}
The magnetic field $\v B$ is gauge invariant, so the only term of
(\ref{eq:F2}) that changes is the last one, and (\ref{eq:dF}) follows.
The boundary form is (\ref{eq:gaugebody}) with $\gl$ replaced by
$\v w\cdot\nabla\gl$. For (b), we note that the values of $\v w\cdot\nabla\gl$
on $\bdry\gO$ can be any smooth function. For instance, $\gl=x_{1}$
and $\v w=\psi\v e_{1}$ give $\v w\cdot\nabla\gl=\psi$.
\end{proof}

This is the main difference from electrostatics, where the force is
gauge invariant for every charge system {[}GMS26a, Prop.~6.1{]}.
There, a gauge transformation changes the energy of a convected free
charge by a constant. Here, using (\ref{eq:Ue}), it changes the energy
of a convected free current element $I\dee{\v l}$ by $-I\dee{\v l_{t}}\cdot\nabla\gl(\dis_{t}(\v y))$,
which changes in the motion. For a closed circuit inside $\gO$ these
changes add up to zero. For a subbody through the boundary of which
free current flows, they do not add up to zero, and the force derived
from the energy of the subbody alone depends on the gauge. The motion
of the points where the current enters and leaves $\gO$ requires
the rest of the circuit to follow. The energy of that part of the
circuit, which lies outside $\gO$ and includes the leads attached
at the contacts, is not included in $U_{\gO}$. Virtual velocities
that vanish on $\bdry\gO$ do not move these points, and for them
the force is gauge invariant. In Section~\ref{sec:forcesystem} we
show what the gauge-dependent term represents.

\textbf{Gauge invariance of the forces on all subbodies.} What if
we require the force to be gauge invariant on every subbody and for
every virtual velocity? Taking $\gl=x_{1}$ and $\v w=\psi\v e_{1}$
in (\ref{eq:dF}), with $\psi$ arbitrary, we get $\int_{\gO}\v j_{f}\cdot\nabla\psi\dV=0$
for every $\gO$ and $\psi$. With $\psi=x_{i}$ this gives $\int_{\gO}j_{f\,i}\dV=0$
for every $\gO$, so $\v j_{f}=0$. Hence, the forces are gauge invariant
on all subbodies if and only if the energy is, i.e., if and only if
there is no free current. In this sense, unlike the case of the energy,
requiring gauge invariance of the forces on all subbodies does not
lead to a weaker condition analogous to weak gauge invariance. Such
a condition is obtained only if we restrict the virtual velocities,
as we now show.

\textbf{The role of conservation of charge.} Conservation of charge
is what makes the force gauge invariant for virtual velocities that
vanish near the boundary. Indeed, if $\v w$ vanishes in a neighborhood
of $\bdry\gO$, then so does $\psi=\v w\cdot\nabla\gl$, and, as in
Remark~\ref{rem:charge}, (\ref{eq:dF}) gives 
\[
\gd F_{\gO}(\v w)=-\int_{\gO}(\v w\cdot\nabla\gl)\,\nabla\cdot\v j_{f}\dV.
\]
Without conservation of charge, this does not vanish in general, and
the force would depend on the gauge even in the interior of $\gO$.
With it, the only gauge dependence left is the boundary term of Proposition~\ref{prop:forcegauge},
due to the free current that crosses $\bdry\gO$.

\textbf{The free current as a dipole distribution.} By Remark~\ref{rem:Mf},
a divergence-free free current may be written as $\v j_{f}=\nabla\cross\v M_{f}$.
If we regard $\v M_{f}$ as a magnetization and embed it as in (\ref{eq:convect}),
then, since the pullback of forms commutes with the curl, $\nabla\cross\v M_{f\,t}=\v j_{f\,t}$
at every $t$, and the force on the dipole distribution $\v M_{f}$
is gauge invariant by Proposition~\ref{prop:forcegauge}(a). But
the dipole distribution $\v M_{f}$ also carries the surface current
$\v M_{f}\cross\v{\nor}$ on $\bdry\gO$, which the free current does
not have, and which depends on the choice of $\v M_{f}$. The force
on it replaces the gauge-dependent boundary term. Thus, the gauge
dependence of the force on the free current is traded for a dependence
on the choice of the potential $\v M_{f}$, and neither choice is
determined by the current system. Only when the free current does
not cross $\bdry\gO$ can both be avoided.

\subsection{Interpretation as a force system}

\label{sec:forcesystem}

Expression (\ref{eq:F1}) is a linear functional of the first jet
$(\v w,\nabla\v w)$, so it defines a force system in the sense of
Section~\ref{sec:stress}, represented by the variational stress
\begin{equation}
r_{k}=j_{f\,i}A_{i,k}+M_{i}B_{i,k},\qquad\gs_{ij}=A_{i}j_{f\,j}+(\v M\cdot\v B)\gd_{ij}-M_{i}B_{j}.\label{eq:Stot}
\end{equation}
All of it, including the gauge-dependent part, is represented by a
variational stress. It is convenient, however, to separate the terms
that involve $\v A$ itself from those that involve only $\v B$,
as in (\ref{eq:F2}). We write 
\[
F_{\gO}=F^{m}_{\gO}+F^{g}_{\gO},
\]
where $F^{m}_{\gO}$ consists of the first two integrals of (\ref{eq:F2})
and $F^{g}_{\gO}$ is the last one.

\textbf{The gauge-invariant part.} The variational stress of $F^{m}_{\gO}$
is given by 
\begin{equation}
\v r_{m}=\v j_{f}\cross\v B+(\nabla\v B)\trps\v M,\qquad\v{\gs}_{m}=(\v M\cdot\v B)\v I-\v M\otimes\v B,\label{eq:Sm}
\end{equation}
that is, $(\gs_{m})_{ij}=(\v M\cdot\v B)\gd_{ij}-M_{i}B_{j}$. To
find the body force $\v b_{m}=\v r_{m}-\nabla\cdot\v{\gs}_{m}$, we
compute the divergence of $\v{\gs}_{m}$. Since $\nabla\cdot\v B=0$,
\[
\begin{split}(\nabla\cdot\v{\gs}_{m})_{k} & =\big((\v M\cdot\v B)\gd_{kj}-M_{k}B_{j}\big)_{,j}\\
 & =M_{i,k}B_{i}+M_{i}B_{i,k}-M_{k,j}B_{j}-M_{k}B_{j,j}\\
 & =M_{i,k}B_{i}+M_{i}B_{i,k}-M_{k,j}B_{j}.
\end{split}
\]
The term $M_{i}B_{i,k}$ cancels against the dipole force density
in $\v r_{m}$, so that 
\[
(\v b_{m})_{k}=(\v j_{f}\cross\v B)_{k}+M_{k,j}B_{j}-M_{j,k}B_{j}.
\]
The last two terms are the Lorentz force on the bound current. Indeed,
with $(\nabla\cross\v M)_{i}=\eps_{ipq}M_{q,p}$ and (\ref{eq:epsdelta}),
one has 
\begin{equation}
\big((\nabla\cross\v M)\cross\v B\big)_{k}=\eps_{kij}(\nabla\cross\v M)_{i}B_{j}=\eps_{kij}\eps_{ipq}M_{q,p}B_{j}=M_{k,j}B_{j}-M_{j,k}B_{j}.\label{eq:curlMxB}
\end{equation}
For the surface traction, we use $(\v a\cross\v b)\cross\v c=\v b(\v a\cdot\v c)-\v a(\v b\cdot\v c)$
with $\v a=\v M$, $\v b=\v{\nor}$ and $\v c=\v B$. Altogether,
\begin{equation}
\begin{split}\v b_{m} & =\v r_{m}-\nabla\cdot\v{\gs}_{m}=(\v j_{f}+\nabla\cross\v M)\cross\v B=\v j\cross\v B,\\
\v t_{\gO} & =\v{\gs}_{m}\v{\nor}=(\v M\cdot\v B)\v{\nor}-\v M(\v B\cdot\v{\nor})=(\v M\cross\v{\nor})\cross\v B=\v k_{\gO}\cross\v B.
\end{split}
\label{eq:bm}
\end{equation}
The magnetic field exerts the Lorentz force on the total volume current
$\v j=\v j_{f}+\v j_{b}$, free and bound, and on the bound surface
current exposed on $\bdry\gO$. The two representations of $F^{m}_{\gO}$,
(\ref{eq:F2}) by a variational stress and (\ref{eq:bm}) by a body
force and a traction, are the continuum counterparts of the two computations
of Section~\ref{sec:dipole}\textemdash the first through the magnetic
moment, the second through the individual current elements.

\textbf{The gauge-dependent part.} Since $j_{f\,j}(A_{i}w_{i})_{,j}=j_{f\,j}A_{i,j}w_{i}+A_{i}j_{f\,j}w_{i,j}$,
the variational stress of $F^{g}_{\gO}$ is 
\[
r^{g}_{i}=A_{i,j}j_{f\,j},\qquad\gs^{g}_{ij}=A_{i}j_{f\,j},\quad\text{i.e.,}\quad\v{\gs}^{g}=\v A\otimes\v j_{f}.
\]
Its body force is $r^{g}_{i}-(A_{i}j_{f\,j})_{,j}=-A_{i}\,\nabla\cdot\v j_{f}$,
which vanishes by conservation of charge, and its surface traction
is $\v{\gs}^{g}\v{\nor}=(\v j_{f}\cdot\v{\nor})\v A$. Thus, the gauge-dependent
part does not prevent the representation as a force system. Rather,
it is a force system of its own, with a stress $\v A\otimes\v j_{f}$
carried by the free current. It has no body force, and its traction
acts only where free current crosses $\bdry\gO$. It depends on the
gauge, as it should by Proposition~\ref{prop:forcegauge}. Adding
the two parts, the whole force system (\ref{eq:Stot}) has the body
force and the surface traction 
\begin{equation}
\v b=\v j\cross\v B,\qquad\v t_{\gO}=\v k_{\gO}\cross\v B+(\v j_{f}\cdot\v{\nor})\,\v A,\label{eq:Lorentz}
\end{equation}
and 
\[
F_{\gO}(\v w)=\int_{\gO}(\v j\cross\v B)\cdot\v w\dV+\int_{\bdry\gO}(\v k_{\gO}\cross\v B)\cdot\v w\dA+\int_{\bdry\gO}(\v A\cdot\v w)\,\v j_{f}\cdot\v{\nor}\dA.
\]

\textbf{What the field determines, and what it does not.} By (\ref{eq:Lorentz}),
the body force $\v j\cross\v B$ is gauge invariant, and it is determined
by the field, whether or not free current crosses $\bdry\gO$. Indeed,
if $\v w$ vanishes in a neighborhood of $\bdry\gO$, the boundary
terms drop out, and 
\[
F_{\gO}(\v w)=\int_{\gO}(\v j\cross\v B)\cdot\v w\dV
\]
in every gauge. Such virtual velocities can be concentrated about
any interior point of $\gO$, so they determine the body force density
everywhere. No closed circuit is needed for that. What the field does
not determine is the part $(\v j_{f}\cdot\v{\nor})\v A$ of the traction
on the parts of $\bdry\gO$ through which free current flows. For
a given potential $\v A$, this part is determined by the energy.
However, under a gauge transformation $\v A\mapsto\v A+\nabla\gl$,
which does not change the field, it changes by $(\v j_{f}\cdot\v{\nor})\nabla\gl$,
so it is gauge invariant only where it vanishes, in accordance with
Proposition~\ref{prop:forcegauge}. It is the traction of the stress
$\v A\otimes\v j_{f}$, which has no body force, and we show in Appendix~\ref{app:potmom}
that it is not a force on the matter. In detail, let a subbody $\gO$
be divided by a surface $S$ into two subbodies $\gO_{1}$ and $\gO_{2}$,
and let $\v{\nor}$ be the unit normal to $S$ pointing from $\gO_{1}$
into $\gO_{2}$. On $S$, the traction acting on $\gO_{1}$ is $(\v j_{f}\cdot\v{\nor})\v A$,
and the traction acting on $\gO_{2}$ is $(\v j_{f}\cdot(-\v{\nor}))\v A$.
The two cancel, so that 
\[
F^{g}_{\gO_{1}}(\v w)+F^{g}_{\gO_{2}}(\v w)=F^{g}_{\gO}(\v w)
\]
for every $\v w$. Moreover, if the free current closes within $\gO$,
i.e., $\v j_{f}\cdot\v{\nor}=0$ on $\bdry\gO$, then, since $\nabla\cdot\v j_{f}=0$,
\[
\begin{split}F^{g}_{\gO}(\v w) & =\int_{\gO}\v j_{f}\cdot\nabla(\v A\cdot\v w)\dV\\
 & =\int_{\bdry\gO}(\v A\cdot\v w)\,\v j_{f}\cdot\v{\nor}\dA-\int_{\gO}(\v A\cdot\v w)\,\nabla\cdot\v j_{f}\dV=0
\end{split}
\]
for every $\v w$. Thus, $\v A\otimes\v j_{f}$ is a self-equilibrated
stress, and adding it to a stress changes neither the body force nor
the total force on such a body. This is the familiar static indeterminacy
of continuum mechanics. The loads on a body do not determine the stress
in it. Here, for a given potential, the stress $\v A\otimes\v j_{f}$
is determined, but it changes under gauge transformations, while the
body force and the total force on such a body remain unchanged. Thus,
the field $\v B$ alone does not determine the interactions across
cuts through the free current. Steady currents of bounded support
always close in some region, at worst in all of $\rthree$, through
the leads and the sources. The energy of such a region determines
the total force on it and the body force $\v j\cross\v B$ everywhere
in it.

\textbf{The gauge-dependent part as a flux of potential momentum.}
In Appendix~\ref{app:potmom}, we show that the stress $\v A\otimes\v j_{f}$
is the flux of the potential momentum $q\v A$ carried by the charge
carriers of the free current, and that its traction $(\v j_{f}\cdot\v{\nor})\v A$
is the rate at which potential momentum crosses a surface element.
It is not a force exerted on the matter. Within the wire, the carriers
exchange momentum with the material. Since the current is steady and
the inertia of the carriers is negligible, the force exerted on the
carriers by the material balances the Lorentz force on them, and,
by reaction, the force on the material is the Lorentz force. The appendix
also shows that the stress $\v A\otimes\v j_{f}$ is conjugate to
the deformation gradient in the same way that the potential momentum
is conjugate to the velocity of a carrier, and that $F^{g}_{\gO}$
is the natural boundary term of the corresponding Euler\textendash Lagrange
problem, the counterpart of the endpoint term that Hamilton's principle
discards for a single carrier. For a subbody crossed by free current,
the energy of the subbody is therefore not a potential for the force
on its matter.

\textbf{The force on a subbody.} We conclude that the force exerted
by the magnetic field on the matter in a subbody $\gO$ is the gauge-invariant
part, 
\begin{equation}
F^{m}_{\gO}(\v w)=\int_{\gO}(\v j\cross\v B)\cdot\v w\dV+\int_{\bdry\gO}(\v k_{\gO}\cross\v B)\cdot\v w\dA,\label{eq:Fmatter}
\end{equation}
that is, the Lorentz force on the total current, free and bound, and
on the bound surface current. It is represented by the variational
stress $(\v r_{m},\v{\gs}_{m})$ of (\ref{eq:Sm}), it is gauge invariant,
and it is defined for every subbody, whether or not free current crosses
its boundary. It agrees with the force derived from the energy for
every body around which the free current closes, and, for every subbody,
for virtual velocities that vanish near its boundary.

\textbf{Contrast between the two levels.} At the level of potentials,
gauge invariance of the energy characterizes $\v j_{f}=0$. At the
mechanical level, the zeroth-order term $\v r_{m}$ is in general
nonzero even for a system without free current, and $F^{m}_{\gO}(\v c)=\v c\cdot\int_{\gO}\v r_{m}\dV\neq0$.
The magnetic field does exert a net force on matter, and this force
must be balanced by other forces, e.g., by the mechanical stress in
the material or by external supports.

\textbf{The active stress and the couple density.} The tensor $\v{\gs}_{m}=(\v M\cdot\v B)\v I-\v M\otimes\v B$
acts on the velocity gradient, so it is a stress. It is an active
stress exerted by the magnetic field. It is neither the stress in
the material nor the Maxwell stress, and it is not symmetric. Its
isotropic part $(\v M\cdot\v B)\v I$ is a pressure-like term that
comes from the convection of the dipoles as area elements, through
the factor $J_{t}$ in (\ref{eq:mt}). It has no counterpart in electrostatics,
where the dipoles are line elements and the active stress is $\v E\otimes\v P$.
For a rigid rotation $\v w=\v{\go}\cross\v x$, as in (\ref{eq:couple-m}),
\begin{equation}
(\gs_{m})_{ij}w_{i,j}=\v{\go}\cdot(\v M\cross\v B),\label{eq:couple}
\end{equation}
so the skew-symmetric part of $\v{\gs}_{m}$ is the couple density
$\v M\cross\v B$, the continuum counterpart of the couple $\v m\cross\v B$
of Section~\ref{sec:dipole}. The dipole force density $(\nabla\v B)\trps\v M$
is the counterpart of $(\v P\cdot\nabla)\v E$ in electrostatics.
Since $\v B$ is not a gradient, it differs in general from $(\v M\cdot\nabla)\v B$.
By (\ref{eq:curlMxB}) with the roles of the fields exchanged, the
difference is $\v M\cross(\nabla\cross\v B)$.

\section{Examples}

\label{sec:examples}

In the examples, $\v B$ is the external field, and the free current
is divergence free and does not cross $\bdry\gO$, unless stated otherwise,
so that the last term of (\ref{eq:F2}) vanishes.
\begin{example}[Total force and torque]
\label{ex:total} For a uniform virtual velocity $\v w=\v c$, $\nabla\v w=0$,
and (\ref{eq:F2}) and (\ref{eq:bm}) give $F_{\gO}(\v c)=\v c\cdot\v F$,
with the total force 
\begin{equation}
\v F=\int_{\gO}\big(\v j_{f}\cross\v B+(\nabla\v B)\trps\v M\big)\dV=\int_{\gO}\v j\cross\v B\dV+\int_{\bdry\gO}\v k_{\gO}\cross\v B\dA.\label{eq:Ftot}
\end{equation}
For a rigid rotation about the origin, $\v w=\v{\go}\cross\v x$,
we have $\v r_{m}\cdot(\v{\go}\cross\v x)=\v{\go}\cdot(\v x\cross\v r_{m})$,
and, by (\ref{eq:couple}), $F_{\gO}(\v{\go}\cross\v x)=\v{\go}\cdot\v T$,
where $\v T$ is the total torque given by 
\begin{equation}
\v T=\int_{\gO}\Big[\v x\cross\big(\v j_{f}\cross\v B+(\nabla\v B)\trps\v M\big)+\v M\cross\v B\Big]\dV.\label{eq:Ttot}
\end{equation}
\end{example}

\begin{example}[Uniform field]
\label{ex:uniform} Let $\v B=\v B_{0}$ be uniform. Then $\nabla\v B=0$,
so the dipole force density vanishes, and (\ref{eq:Ftot}) and (\ref{eq:Ttot})
reduce to 
\[
\v F=\Big(\int_{\gO}\v j_{f}\dV\Big)\cross\v B_{0},
\]
\[
\v T=\int_{\gO}\v x\cross(\v j_{f}\cross\v B_{0})\dV+\Big(\int_{\gO}\v M\dV\Big)\cross\v B_{0}.
\]
We now use the assumption that the free current is divergence free
and does not cross $\bdry\gO$. Then, by (\ref{eq:gaugebody}), $\int_{\gO}\v j_{f}\cdot\nabla\gl\dV=0$
for every $\gl$. With $\gl=x_{i}$ this gives $\int_{\gO}j_{f\,i}\dV=0$,
so $\v F=0$. With $\gl=x_{i}x_{j}$, for which $\v j_{f}\cdot\nabla\gl=j_{f\,i}x_{j}+x_{i}j_{f\,j}$,
it gives 
\[
\int_{\gO}x_{j}j_{f\,i}\dV=-\int_{\gO}x_{i}j_{f\,j}\dV,
\]
so the first moment of the free current is skew-symmetric. In particular,
$\int_{\gO}\v x\cdot\v j_{f}\dV=0$. For the torque, we use $\v x\cross(\v j_{f}\cross\v B_{0})=\v j_{f}(\v x\cdot\v B_{0})-\v B_{0}(\v x\cdot\v j_{f})$.
The second term integrates to zero, and, by the skew-symmetry, 
\[
\int_{\gO}j_{f\,k}x_{n}B_{0n}\dV=\half B_{0n}\int_{\gO}\big(x_{n}j_{f\,k}-x_{k}j_{f\,n}\big)\dV.
\]
On the other hand, with $\v a:=\half\int_{\gO}\v x\cross\v j_{f}\dV$,
i.e., $a_{p}=\half\eps_{pqs}\int_{\gO}x_{q}j_{f\,s}\dV$, and (\ref{eq:epsdelta}),
one has 
\[
\begin{split}(\v a\cross\v B_{0})_{k} & =\eps_{kpn}a_{p}B_{0n}=\half\eps_{pnk}\eps_{pqs}B_{0n}\int_{\gO}x_{q}j_{f\,s}\dV\\
 & =\half B_{0n}\int_{\gO}\big(x_{n}j_{f\,k}-x_{k}j_{f\,n}\big)\dV.
\end{split}
\]
It follows that 
\[
\int_{\gO}\v x\cross(\v j_{f}\cross\v B_{0})\dV=\Big(\half\int_{\gO}\v x\cross\v j_{f}\dV\Big)\cross\v B_{0},
\]
and the two contributions to the torque combine: 
\begin{equation}
\begin{split}\v F & =0,\\
\v T & =\Big(\half\int_{\gO}\v x\cross\v j_{f}\dV+\int_{\gO}\v M\dV\Big)\cross\v B_{0}=\v m_{\gO}\cross\v B_{0},
\end{split}
\end{equation}
with the magnetic moment $\v m_{\gO}$ of (\ref{eq:mOmega}), which
includes the moment $\int_{\gO}\v M\dV$ of the magnetization. Since
the net free current vanishes, $\v m_{\gO}$ does not depend on the
point $\v x_{0}$ in (\ref{eq:mOmega}), and we took $\v x_{0}=0$.
This is the continuum counterpart of the vanishing force and the couple
$\v m\cross\v B$ on a single dipole in a uniform field.
\end{example}

\begin{example}[Uniformly magnetized body in a field with uniform gradient]
\label{ex:sphere} Let $\v j_{f}=0$, $\v M=\v M_{0}$ uniform in
$\gO$, and $\v B(\v x)=\v B_{0}+\v G\v x$ with $\v G$ a constant
matrix, so that $B_{i,k}=G_{ik}$. We assume that $\v B$ is the field
of sources outside $\gO$, so that $\nabla\cross\v B=0$ in $\gO$.
The reason for this assumption is that, in magnetostatics, $\nabla\cross\v B=\mu_{0}\v j_{\mathrm{tot}}$,
where $\v j_{\mathrm{tot}}$ is the total current density, free and
bound, that produces $\v B$, and the currents that produce the external
field are outside $\gO$. Since $\nabla\cdot\v B=G_{kk}$, the matrix
$\v G$ is traceless, and since $(\nabla\cross\v B)_{i}=\eps_{ijk}G_{kj}$,
the matrix $\v G$ is symmetric. Since $\v M_{0}$ is uniform, $\v j_{b}=\nabla\cross\v M_{0}=0$,
and the whole bound current is on the surface, $\v k_{\gO}=\v M_{0}\cross\v{\nor}$.
By (\ref{eq:Ftot}), $F_{k}=\int_{\gO}M_{0i}G_{ik}\dV$, i.e., 
\begin{equation}
\v F=|\gO|\,\v G\trps\v M_{0}=\nabla(\v m\cdot\v B),\qquad\v m=|\gO|\,\v M_{0},
\end{equation}
where $|\gO|$ is the volume of $\gO$, and the gradient is taken
at a fixed $\v m$, so that $(\v m\cdot\v B)_{,k}=m_{i}G_{ik}$. This
is the force on the total magnetic moment. For a uniformly magnetized
ball, it is the force on the equivalent point dipole at its center.
In the representation (\ref{eq:bm}) it is the Lorentz force on the
surface current. The bound current of a uniform magnetization is all
on the surface, and so is the force on it.
\end{example}

\begin{example}[A smoothly tapered magnetization]
\label{ex:taper} Let $\v B=k(x_{2},x_{1},0)$, which is divergence
and curl free, and let $\v j_{f}=0$ and $\v M=f(\v x)\v e_{1}$,
where $f\geq0$ is smooth and compactly supported in the interior
of $\gO$. Then $\v k_{\gO}=\v M\cross\v{\nor}=0$ on $\bdry\gO$,
and by (\ref{eq:Ftot}), with $B_{1}=kx_{2}$, 
\begin{equation}
F_{k}=\int_{\gO}M_{1}B_{1,k}\dV,\qquad\v F=k\Big(\int_{\gO}f\dV\Big)\v e_{2},
\end{equation}
the attraction of the dipoles toward the region where $\v M\cdot\v B=kfx_{2}$
is larger, for $k>0$. There is no surface current, and in the representation
(\ref{eq:bm}) the whole force acts on the bound volume current $\nabla\cross\v M=(0,f_{,3},-f_{,2})$.
\end{example}

\begin{example}[A wire crossing the subbody]
\label{ex:wire} Let $\v B=B_{0}\v e_{3}$ be uniform, and let a
straight wire carrying the current $I$ in the direction $\v e_{1}$
cross $\gO$ along the segment from $\v y_{1}=(-a,0,0)$ to $\v y_{2}=(a,0,0)$,
so that $\v j_{f}=I\,\v e_{1}\gd_{L}$, where $\gd_{L}$ is the line
measure on the segment. It is further assumed that $\v M=0$. Here
the free current crosses $\bdry\gO$ at the two contacts $\v y_{1}$
and $\v y_{2}$. The Lorentz force on the segment is $2aI\,\v e_{1}\cross B_{0}\v e_{3}=-2aIB_{0}\,\v e_{2}$.
The last term of (\ref{eq:F2}) is 
\begin{equation}
\begin{split}\int_{\gO}\v j_{f}\cdot\nabla(\v A\cdot\v w)\dV & =I\int^{a}_{-a}\frac{\bdry}{\bdry x_{1}}(\v A\cdot\v w)(x_{1},0,0)\,\dee{x_{1}}\\
 & =I\big[(\v A\cdot\v w)(\v y_{2})-(\v A\cdot\v w)(\v y_{1})\big].
\end{split}
\end{equation}
For a uniform velocity $\v w=\v c$, the force derived from the energy
is therefore $F_{2}=-2aIB_{0}+I\big(A_{2}(\v y_{2})-A_{2}(\v y_{1})\big)$.
Consider three potentials of the same field: $\v A=-B_{0}x_{2}\v e_{1}$,
$\v A=\half B_{0}(-x_{2},x_{1},0)$ and $\v A=B_{0}x_{1}\v e_{2}$.
Each has curl $B_{0}\v e_{3}$, and they differ by gauge fields. They
give $A_{2}(\v y_{2})-A_{2}(\v y_{1})=0$, $aB_{0}$ and $2aB_{0}$,
respectively, so that $F_{2}$ is $-2aIB_{0}$, $-aIB_{0}$ and $0$.

The following discussion proposes an explanation of these results.
Firstly, the results confirm Proposition~\ref{prop:forcegauge}.
The last term of (\ref{eq:F2}), although it was derived from the
potential energy of $\gO$, is not gauge invariant when the free current
crosses $\bdry\gO$, and the three gauges give three different values.
Secondly, by Appendix~\ref{app:potmom}, this term is the net rate
at which the charge carriers transport potential momentum out of $\gO$.
Here, the carriers enter at $\v y_{1}$ and leave at $\v y_{2}$,
so that potential momentum leaves $\gO$ at the net rate $I\big(\v A(\v y_{2})-\v A(\v y_{1})\big)$,
and its component along $\v c$ is the gauge term above. The three
gauges assign different potential momenta to the carriers at the contacts.
For the gauge $\v A=-B_{0}x_{2}\v e_{1}$, $\v A$ vanishes on the
$x_{1}$-axis, the carriers on the wire have no potential momentum,
and the energy gives exactly the Lorentz force. Thirdly, the Lorentz
force $-2aIB_{0}\,\v e_{2}$ is the same in all three gauges, and,
by (\ref{eq:Fmatter}), it is the force exerted by the field on the
wire. The energy of $\gO$ does determine it for virtual velocities
that vanish at the contacts. For the closed circuit as a whole, the
free current does not cross the boundary, and the force derived from
the energy is the Lorentz force for every virtual velocity.
\end{example}

\section{Concluding remarks}

\label{sec:concl}

\textbf{Summary.} Magnetostatics is the stress theory we get by replacing
velocities with vector potentials. Cauchy's theorem represents every
current system by a pair $(\v r,\v{\Sigma})$, the analog of a variational
stress. Gauge transformations assume the role of rigid motions, and
weak gauge invariance, the assumption that a gauge transformation
changes the energy only through the values of $\nabla\gl$, is the
analog of the symmetry of the stress. It holds if and only if $\v{\Sigma}$
is skew-symmetric. It follows that $\v{\Sigma}$ is represented by
the magnetization $\v M$, and the current splits uniquely into free
current $\v j_{f}$ and a dipole distribution carrying the bound currents
$\nabla\cross\v M$ and $\v M\cross\v{\nor}$. Gauge invariance of
the energy is not a postulate. It plays the same role as the balance
of forces on every subbody in mechanics. It holds if and only if no
subbody carries a net current, if and only if $\v j_{f}=0$, and if
and only if the energy is $-\int\v M\cdot\v B\dV$. Between the two,
invariance under gauge transformations that vanish near the boundary
is equivalent to conservation of charge. The mechanical forces, viewed
as collections of linear functionals on material velocity fields,
one for each subbody, are derived for currents embedded in the body.
They consist of the Lorentz force on the total current, free and bound,
or, equivalently, of the force $\v j_{f}\cross\v B$ on the free current,
the dipole force density $(\nabla\v B)\trps\v M$, and the asymmetric
active stress $(\v M\cdot\v B)\v I-\v M\otimes\v B$, whose skew-symmetric
part is the couple density $\v M\cross\v B$. In addition, the force
derived from the energy contains a gauge-dependent term where free
current crosses the boundary of the subbody. It is the flux of potential
momentum carried by the charge carriers, and it is not a force on
the matter.

The equations of magnetostatics in a magnetizable medium are (e.g.,
{[}Jac99, Zan13{]}) 
\begin{equation}
\nabla\cdot\v B=0,\qquad\nabla\cross\v H=\v j_{f},\qquad\v H=\frac{1}{\mu_{0}}\v B-\v M,\label{eq:maxwell}
\end{equation}
where $\v H$ is the magnetic field intensity, $\v j_{f}$ the free
current density and $\v M$ the magnetization. We have not used any
of these equations. We have treated the potential $\v A$ and the
current system $(\v j_{f},\v M)$ as independent, just as, in stress
theory, a velocity field is applied to a fixed force system regardless
of the constitutive properties of the body. Maxwell's equations come
in once we require the potential to be the one generated by the currents.
Note that the second equation implies $\nabla\cdot\v j_{f}=0$. The
field equations require charge conservation, in agreement with Remark~\ref{rem:charge}.

Since $\v B=\nabla\cross\v A$, the first equation of (\ref{eq:maxwell})
holds identically. As in electrostatics, it is the analog of the compatibility
of a strain field derived from a displacement field. We used it in
the derivation of the Lorentz force on the dipoles in Section~\ref{sec:dipole}
and in (\ref{eq:bm}).

\textbf{Generalization.} Electrostatics {[}GMS26a{]} and magnetostatics
are special cases of a theory of electrodynamics with $p$-form potentials
$\ga$ on an $n$-dimensional manifold. Electrostatics is the case
$n=3$, $p=0$, magnetostatics is the case $n=3$, $p=1$, and electromagnetism
is the case $n=4$, $p=1$, in spacetime. Under the analog of weak
gauge invariance, the variational stress takes the form $\mathfrak{J}_{f}\wedge\ga+(-1)^{n-p-1}\mathfrak{g}\wedge\dee{\ga}$,
where $\mathfrak{J}_{f}$ is a free source and $\mathfrak{g}$ is
an $(n-p-1)$-form generalizing the magnetization. Gauge invariance
of the energy characterizes the case $\mathfrak{J}_{f}=0$. In magnetostatics,
$\mathfrak{J}_{f}=-\v j_{f}\mathbin{\lrcorner}\dV$, and $\mathfrak{g}$
is the 1-form of $\v M$. We treat the general case in {[}GMS26b{]}.

\section*{Acknowledgment}

The authors are grateful to Claude (Anthropic), an AI system, which
contributed substantially to the development of the arguments, to
the derivations, and to the preparation of this manuscript, and to
the many colleagues with whom earlier versions of the manuscript were
discussed.

\appendix

\section{Potential momentum and the gauge-dependent part of the force}

\label{app:potmom}In this appendix, we interpret the gauge-dependent
part $F^{g}_{\gO}$ of the force derived from the energy in Section~\ref{sec:forcesystem}.
Since the magnetization does not contribute to $F^{g}_{\gO}$, we
consider only the free current. We assume that the potential $\v A$
is given and does not depend on time, and that $\nabla\cdot\v j_{f}=0$.

\textbf{A single charge carrier.} Consider a single charge carrier,
with charge $q$ and velocity $\v u$, moving along the wire through
the field of the potential $\v A$. We recall that, in the nonrelativistic
case, the Lagrangian of a charge $q$ of mass $m$ moving with velocity
$\v u$ in the field of the potential $\v A$ is $L=\half m|\v u|^{2}+q\,\v u\cdot\v A$,
so that $-q\,\v u\cdot\v A$ is a velocity-dependent potential {[}GPS02,
Sect.~1.5{]}. The canonical momentum is $\bdry L/\bdry\v u=m\v u+q\v A$,
and the term $q\v A$ is the potential momentum. The Euler\textendash Lagrange
equations, 
\[
\frac{\dee{}}{\dee t}(mu_{k}+qA_{k})=\frac{\bdry L}{\bdry x_{k}}=q\,u_{i}A_{i,k},
\]
hold with $\v x$ and $\v u$ as independent variables. Since the
potential $\v A$ does not depend on time, the potential momentum
changes along the path of the carrier at the rate $\dee (qA_{k})/\dee t=q\,u_{i}A_{k,i}$.
Thus, 
\[
m\dot{u}_{k}=q\,u_{i}A_{i,k}-q\,u_{i}A_{k,i},
\]
and, by (\ref{eq:dlxB}), the right-hand side is the component $k$
of the Lorentz force $q\,\v u\cross\v B$. Its first term is minus
the derivative of the velocity-dependent potential $-q\,\v u\cdot\v A$
with respect to position. The second is minus the rate of change of
the potential momentum of the carrier. The second term results from
an integration by parts in time. In Hamilton's principle, the variation
of the action $\int L\,\dee t$ for a variation $\boldsymbol{\eta}$
of the path contains the term 
\[
\int^{t_{2}}_{t_{1}}q\v A\cdot\dot{\boldsymbol{\eta}}\,\dee t=\Big[q\v A\cdot\boldsymbol{\eta}\Big]^{t_{2}}_{t_{1}}-\int^{t_{2}}_{t_{1}}\frac{\dee (q\v A)}{\dee t}\cdot\boldsymbol{\eta}\,\dee t,
\]
and the endpoint term is discarded because $\boldsymbol{\eta}$ vanishes
at the initial and final times.

\textbf{The continuum.} For the continuum, we use only the free current
density $\v j_{f}$, without reference to the charge carriers or to
their charge density. A configuration of the matter is a mapping $\chi\from\gO\to\rthree$,
with the deformation gradient $\v F=\nabla\chi$, where $\gO$ is
the region occupied by the matter in the reference configuration.
We use capital indices for components with respect to coordinates
$x_{I}$ in the reference configuration. As in (\ref{eq:convect}),
the current is convected with the matter, so that a current element
$\v j_{f}\dV$ is carried to $\v F\v j_{f}\dV$. Since $\v F\v j_{f}=(\v j_{f}\cdot\nabla)\chi$,
the current element is the derivative of $\chi$ along the current
lines. As for the Lagrangian of a single carrier, we define the density
$\ell$ as a function of independent variables,
\begin{equation}
\ell(\v x,\v y,\v H)=\v A(\v y)\cdot\v H\v j_{f}(\v x)=A_{k}(\v y)H_{kI}\,j_{f\,I}(\v x),\label{eq:ell}
\end{equation}
where $\v x\in\gO$, $\v y\in\rthree$ is a point in space, and $\v H$
is a $3\times3$ matrix. The energy of the configuration $\chi$ is
\[
U[\chi]=-\int_{\gO}\ell\big(\v x,\chi(\v x),\v F(\v x)\big)\dV=-\int_{\gO}(\v A\comp\chi)\cdot\v F\v j_{f}\dV.
\]
There is no counterpart of the kinetic energy, since the motion of
the carriers relative to the matter is not a degree of freedom. The
partial derivatives of $\ell$ are
\begin{equation}
\frac{\bdry\ell}{\bdry y_{k}}=A_{i,k}(\v y)H_{iI}\,j_{f\,I}(\v x),\qquad P_{kI}:=\frac{\bdry\ell}{\bdry H_{kI}}=A_{k}(\v y)\,j_{f\,I}(\v x).\label{eq:Pconj}
\end{equation}
For a variation of the configuration, given by a vector field $\boldsymbol{\eta}$
on $\gO$,
\[
\begin{split}\gd\int_{\gO}\ell\dV & :=\frac{\dee{}}{\dee{\eps}}\int_{\gO}\ell\big(\v x,\chi+\eps\boldsymbol{\eta},\v F+\eps\nabla\boldsymbol{\eta}\big)\dV\Big|_{\eps=0}\\
 & =\int_{\gO}\Big(\frac{\bdry\ell}{\bdry y_{k}}\eta_{k}+P_{kI}\eta_{k,I}\Big)\dV,
\end{split}
\]
where the partial derivatives are evaluated at $\v y=\chi(\v x)$
and $\v H=\v F(\v x)$.

To integrate the second term by parts, we compute the divergence of
$P_{kI}=(A_{k}\comp\chi)\,j_{f\,I}$. By the chain rule, $(A_{k}\comp\chi)_{,I}=(A_{k,i}\comp\chi)F_{iI}$,
and, since $\nabla\cdot\v j_{f}=j_{f\,I,I}=0$,
\[
\begin{split}P_{kI,I} & =(A_{k}\comp\chi)_{,I}\,j_{f\,I}+(A_{k}\comp\chi)\,j_{f\,I,I}\\
 & =(A_{k,i}\comp\chi)F_{iI}\,j_{f\,I}=(A_{k,i}\comp\chi)(\v F\v j_{f})_{i}.
\end{split}
\]
By the divergence theorem,
\[
\int_{\gO}P_{kI}\eta_{k,I}\dV=\int_{\bdry\gO}P_{kI}\nu_{I}\eta_{k}\dA-\int_{\gO}P_{kI,I}\eta_{k}\dV,
\]
where $P_{kI}\nu_{I}\eta_{k}=\big((\v A\comp\chi)\cdot\boldsymbol{\eta}\big)(\v j_{f}\cdot\v{\nor})$.
Substituting the expressions for $\bdry\ell/\bdry y_{k}$ and $P_{kI,I}$,
we obtain
\[
\begin{split}\gd\int_{\gO}\ell\dV & =\int_{\gO}(\v F\v j_{f})_{i}\,\big((A_{i,k}-A_{k,i})\comp\chi\big)\,\eta_{k}\dV\\
 & \qquad+\int_{\bdry\gO}\big((\v A\comp\chi)\cdot\boldsymbol{\eta}\big)(\v j_{f}\cdot\v{\nor})\dA.
\end{split}
\]
At the reference configuration, $\chi$ is the identity and $\v F=\v I$,
and, for $\boldsymbol{\eta}=\v w$, the variation of $\int_{\gO}\ell\dV=-U$
is the force functional $F_{\gO}(\v w)$. By (\ref{eq:dlxB}), $j_{f\,i}(A_{i,k}-A_{k,i})=(\v j_{f}\cross\v B)_{k}$,
and
\begin{equation}
F_{\gO}(\v w)=\int_{\gO}(\v j_{f}\cross\v B)\cdot\v w\dV+\int_{\bdry\gO}(\v A\cdot\v w)(\v j_{f}\cdot\v{\nor})\dA.\label{eq:Ffree2}
\end{equation}

This derivation is the counterpart of Hamilton's principle for a single
carrier, with the time along the path of the carrier replaced by the
position along the current lines. The current element $\v F\v j_{f}$
plays the role of the velocity, the density $\ell$, a function of
the independent variables $\v y$ and $\v H$, plays the role of the
term $q\,\v u\cdot\v A$ of the Lagrangian, a function of the independent
variables $\v x$ and $\v u$, and the tensor $\v P$ of (\ref{eq:Pconj}),
which at the reference configuration is the stress $\v A\otimes\v j_{f}$
of the gauge-dependent part, plays the role of the potential momentum
$\bdry L/\bdry\v u=q\v A$. Thus, the stress $\v A\otimes\v j_{f}$
is conjugate to the deformation gradient in the same way that the
potential momentum is conjugate to the velocity. The divergence of
$\v P$ replaces the time derivative of the potential momentum, and
the Euler\textendash Lagrange expression $\bdry\ell/\bdry\v y-\mathrm{Div}\,\v P$
is the Lorentz force density, as the Euler\textendash Lagrange equations
give the Lorentz force on the carrier. The boundary term in (\ref{eq:Ffree2})
is the gauge-dependent part $F^{g}_{\gO}(\v w)$, with the traction
$(\v j_{f}\cdot\v{\nor})\v A$, and it is the counterpart of the endpoint
term of Hamilton's principle. For a single carrier, the endpoint term
is discarded because the variation vanishes at the initial and final
times. For the continuum, the variation does not vanish on $\bdry\gO$,
and the boundary term remains wherever the current lines cross $\bdry\gO$.
If the free current closes within $\gO$, then $\v j_{f}\cdot\v{\nor}=0$
on $\bdry\gO$, and the boundary term vanishes. Gauge transformations
fit the same pattern. For a single carrier, a gauge transformation
$\v A\mapsto\v A+\nabla\gl$ adds to the Lagrangian the total time
derivative $q\,\dee{\gl}/\dee t$, which changes the action only by
endpoint terms and does not change the equations of motion. For the
continuum, it adds to $\ell$ the divergence $(\gl_{,k}\comp\chi)F_{kI}j_{f\,I}=\big((\gl\comp\chi)\,j_{f\,I}\big)_{,I}$,
which changes only the boundary term and leaves the Lorentz force
density unchanged.

\textbf{The gauge-dependent part as a flux of potential momentum.}
The boundary term in (\ref{eq:Ffree2}) can be interpreted directly.
For a single carrier, since $\v A$ and $\v w$ do not depend on time,
the quantity $q\v A\cdot\v w$, its potential momentum paired with
the virtual velocity at its position, changes along its path at the
rate $q\,u_{i}(A_{k}w_{k})_{,i}$. Summing over the carriers in a
volume element gives the density $j_{f\,i}(A_{k}w_{k})_{,i}=\v j_{f}\cdot\nabla(\v A\cdot\v w)$.
Integrating over $\gO$ and using $\nabla\cdot\v j_{f}=0$, we obtain
the last term of (\ref{eq:F2}) and of (\ref{eq:Ffree2}), 
\[
F^{g}_{\gO}(\v w)=\int_{\gO}\v j_{f}\cdot\nabla(\v A\cdot\v w)\dV=\int_{\bdry\gO}(\v A\cdot\v w)(\v j_{f}\cdot\v{\nor})\dA.
\]
Thus, for every virtual velocity field $\v w$, $F^{g}_{\gO}(\v w)$
is the rate at which the carriers transport potential momentum out
of $\gO$, where the potential momentum crossing each boundary point
is paired with the virtual velocity at that point. For a uniform virtual
velocity $\v w=\v c$, it is $\v c\cdot\int_{\bdry\gO}\v A\,(\v j_{f}\cdot\v{\nor})\dA$,
where the integral is the net rate at which potential momentum leaves
$\gO$. We conclude that the force derived from the energy is the
Lorentz force plus the rate at which potential momentum leaves $\gO$
with the charge carriers. Accordingly, the stress $\v A\otimes\v j_{f}$
is the flux of potential momentum carried by the free current, and
its traction $(\v j_{f}\cdot\v{\nor})\v A$ is the rate at which potential
momentum crosses a surface element. It is not a force exerted on the
matter. The flux of potential momentum depends on the gauge because
the potential momentum $q\v A$ does. Only its circulation around
closed loops, the flux of $\v B$, has a gauge-invariant meaning.
This is also why the term vanishes for a closed circuit, since what
flows out of a region flows into the neighboring one.

In the terminology of analytical mechanics, the magnetic force on
a charge carrier is not conservative, since it is derived from the
velocity-dependent potential $-q\,\v u\cdot\v A$ rather than from
a potential that depends only on position {[}GPS02, Sects.~1.1 and
1.5{]}. For the continuum, by (\ref{eq:Ffree2}), the force derived
from the energy differs from the Lorentz force by $F^{g}_{\gO}$,
the rate at which potential momentum leaves the subbody with the carriers.
For a region in which the free current closes, this term vanishes,
and the force on the matter is obtained from the energy. For a subbody
crossed by free current, the energy of the subbody is not a potential
for the force on its matter.


\begin{thebibliography}{GMS26a}
\bibitem[Bro66]{Bro66} W.~F. Brown, Jr. \emph{Magnetoelastic Interactions}.
Springer Tracts in Natural Philosophy, Vol.~9. Springer, 1966.

\bibitem[DO14]{DO14} L. Dorfmann, R. W. Ogden. \emph{Nonlinear Theory
of Electroelastic and Magnetoelastic Interactions}. Springer, 2014.

\bibitem[EM90]{EM90} A. C. Eringen, G. A. Maugin. \emph{Electrodynamics
of Continua I: Foundations and Solid Media}. Springer, 1990.

\bibitem[GMS26a]{GMS26a} V.~Gol'dshtein, W.~H.~Müller, R.~Segev.
Electrostatics as a stress theory in $\rthree$: free and bound charge,
gauge invariance, and mechanical interactions. arXiv:2608.12900 {[}math-ph{]},
2026.

\bibitem[GMS26b]{GMS26b} V.~Gol'dshtein, W.~H.~Müller, R.~Segev.
$p$-Form electrodynamics as a stress theory: free and bound sources,
gauge invariance, and mechanical interactions. In preparation.

\bibitem[GPS02]{GPS02} H.~Goldstein, C.~Poole, J.~Safko. \emph{Classical
Mechanics}. Addison-Wesley, 3rd ed., 2002.

\bibitem[Jac99]{Jac99} J.~D.~Jackson. \emph{Classical Electrodynamics}.
Wiley, 3rd ed., 1999.

\bibitem[Kov00]{Kov00} Y. Kovetz. \emph{Electromagnetic Theory}.
Oxford University Press, 2000.

\bibitem[Mau88]{Mau88} G. A. Maugin. \emph{Continuum Mechanics of
Electromagnetic Solids}. North-Holland, 1988.

\bibitem[Nol59]{Nol59} W.~Noll. The foundations of classical mechanics
in the light of recent advances in continuum mechanics. In \emph{The
Axiomatic Method, with Special Reference to Geometry and Physics}
(L.~Henkin, P.~Suppes, A.~Tarski, eds.), North-Holland, 1959, 266\textendash 281.

\bibitem[PP62]{PP62} W.~K.~H.~Panofsky, M.~Phillips. \emph{Classical
Electricity and Magnetism}. Addison-Wesley, 2nd ed., 1962.

\bibitem[Seg02]{Seg02} R.~Segev. Metric-independent analysis of
the stress-energy tensor. \emph{J. Math. Phys.} 43 (2002), 3220\textendash 3231.

\bibitem[Seg23]{Seg23} R.~Segev. \emph{Foundations of Geometric
Continuum Mechanics}. Birkhäuser, 2023.

\bibitem[TT60]{TT60} C.~Truesdell, R.~A.~Toupin. The classical
field theories. In \emph{Handbuch der Physik}, Vol.~III/1 (S.~Flügge,
ed.), Springer, 1960, 226\textendash 793.

\bibitem[Zan13]{Zan13} A.~Zangwill. \emph{Modern Electrodynamics}.
Cambridge University Press, 2013.

\end{thebibliography}
\end{document}